\documentclass[12pt, draftclsnofoot, onecolumn]{IEEEtran}
\usepackage{amssymb}
\usepackage{amsmath}
\usepackage{cite}
\usepackage{url}
\usepackage{xcolor}
\usepackage{cite,graphicx,amsmath,amssymb}
\usepackage{subfigure}
\usepackage{citesort}
\usepackage{fancyhdr}
\usepackage{mdwmath}
\usepackage{mdwtab}
\usepackage{caption}
\usepackage{amsthm}
\usepackage{setspace}
\usepackage{algorithm}
\usepackage{algorithmic}
\usepackage{makecell}
\usepackage{diagbox}
\newcommand{\bm}[1]{\mbox{\boldmath{$#1$}}}

\newtheorem{remark}{Remark}
\newtheorem{theorem}{Theorem}

\newtheorem{lemma}{Lemma}

\newtheorem{corollary}{Corollary}

\allowdisplaybreaks
\makeatletter
\def\ScaleIfNeeded{%
\ifdim\Gin@nat@width>\linewidth \linewidth \else \Gin@nat@width
\fi } \makeatother

\begin{document}
\title{Joint Deployment and Multiple Access Design for Intelligent Reflecting Surface Assisted Networks}
%

\author{

Xidong~Mu,~\IEEEmembership{Student Member,~IEEE,}
        Yuanwei~Liu,~\IEEEmembership{Senior Member,~IEEE,}
       Li~Guo,~\IEEEmembership{Member,~IEEE,}
       Jiaru~Lin,~\IEEEmembership{Member,~IEEE,}
       and Robert~Schober,~\IEEEmembership{Fellow,~IEEE}

\thanks{X. Mu, L. Guo, and J. Lin are with Beijing University of Posts and Telecommunications, Beijing,
China. (email:\{muxidong, guoli, jrlin\}@bupt.edu.cn).}
\thanks{Y. Liu is with Queen Mary University of London, London, UK. (email:yuanwei.liu@qmul.ac.uk).}
\thanks{R. Schober is with the Institute for Digital Communications, Friedrich-Alexander-University Erlangen-N{\"u}rnberg (FAU), Germany (e-mail: robert.schober@fau.de).}
}

\maketitle
\vspace{-1.2cm}
\begin{abstract}
\vspace{-0.5cm}
The fundamental intelligent reflecting surface (IRS) deployment problem is investigated for IRS-assisted networks, where one IRS is arranged to be deployed in a specific region for assisting the communication between an access point (AP) and multiple users. Specifically, three multiple access schemes are considered, namely non-orthogonal multiple access (NOMA), frequency division multiple access (FDMA), and time division multiple access (TDMA). The weighted sum rate maximization problem for joint optimization of the deployment location and the reflection coefficients of the IRS as well as the power allocation at the AP is formulated. The non-convex optimization problems obtained for NOMA and FDMA are solved by employing monotonic optimization and semidefinite relaxation to find a performance upper bound. The problem obtained for TDMA is optimally solved by leveraging the \emph{time-selective} nature of the IRS. Furthermore, for all three multiple access schemes, low-complexity suboptimal algorithms are developed by exploiting alternating optimization and successive convex approximation techniques, where a \emph{local region optimization} method is applied for optimizing the IRS deployment location. Numerical results are provided to show that: 1) near-optimal performance can be achieved by the proposed suboptimal algorithms; 2) \emph{asymmetric} and \emph{symmetric} IRS deployment strategies are preferable for NOMA and FDMA/TDMA, respectively; 3) the performance gain achieved with IRS can be significantly improved by optimizing the deployment location.
\end{abstract}
\section{Introduction}
\vspace{-0.3cm}
Intelligent reflecting surfaces (IRSs), also known as reconfigurable intelligent surfaces (RISs)~\cite{Basar} and large intelligent surfaces (LISs)~\cite{LiangLISA}, have recently attracted extensive attention from both academia and industry. Generally, an IRS is a planar meta-surface comprising a large number of low-cost passive reflecting elements, which is capable of passively reflecting the incident signals and reconfiguring their amplitudes and phase shifts\cite{di2020smart,WuTowards}. Therefore, by deploying an IRS, the wireless environment becomes controllable. Compared to conventional relaying technologies such as amplify-and-forward (AF) \textcolor{black}{and decode-and-forward (DF) relays}, IRSs consume much less energy since they do not have to be equipped with costly active transmit radio frequency (RF) chains. Additionally, IRSs do not suffer from the self-interference problem due to their nearly passive full-duplex mode of operation. \textcolor{black}{Therefore, IRSs have been envisioned as a promising technology for future sixth generation (6G) communication networks~\cite{Zhao_survey,Huang_Holographic}.}\\
\indent For the integration of IRSs into future wireless networks, multiple access (MA) techniques are essential. Current MA techniques can be loosely classified into two categories, namely orthogonal multiple access (OMA) and non-orthogonal multiple access (NOMA). The key difference between these two MA techniques is whether one resource block (in time, frequency, or code) can serve multiple users or only one user. For downlink NOMA transmission, superposition coding (SC) and successive interference cancellation (SIC) techniques are invoked at the transmitter and receiver, respectively~\cite{Liu2017}. By employing SIC, stronger users are capable of removing the co-channel interference caused by weaker users, before decoding their own signal. In conventional NOMA, the decoding order is determined by the users' channel power gains. With the aid of IRSs, however, the users' decoding order can be designed more freely by reconfiguring the IRS reflection coefficients and optimizing the IRS deployment, which introduces new degrees-of-freedom (DoFs) for IRS-assisted NOMA networks to improve wireless communication performance.
\vspace{-0.6cm}
\subsection{Prior Works}
\vspace{-0.2cm}
\subsubsection{Reflection Coefficient Design in IRS-assisted Networks}
The joint optimization of the IRS reflection coefficients and resource allocation has received considerable research interest as it can significantly improve the performance of IRS-assisted communications~\cite{Wu2019IRS,Huang_Discrete,Huang_EE,Yu,Huang_DRL}. By adjusting the IRS reflection coefficients, the reflected signal can be combined coherently to improve the received signal strength or destructively to mitigate interference. For instance, the authors of~\cite{Wu2019IRS} studied the transmit power minimization problem by jointly designing the transmit beamforming at the access point (AP) and the reflection coefficients at the IRS in both single-user and multi-user scenarios. \textcolor{black}{Based on an IRS element power consumption model, the energy efficiency of IRS-assisted multi-user multiple-input single-output (MISO) networks was maximized by optimizing discrete and continuous IRS reflection coefficients in~\cite{Huang_Discrete} and~\cite{Huang_EE}, respectively.} The authors of~\cite{Yu} maximized the achievable spectral efficiency in a single-user IRS-assisted MISO communication system, where fixed point iteration and manifold optimization techniques were applied for reflection coefficient design. \textcolor{black}{Instead of employing conventional mathematical tools, the authors of \cite{Huang_DRL} optimized the IRS reflection coefficients by invoking deep reinforcement learning methods.} 
\subsubsection{NOMA in IRS-assisted Networks}
The potential performance gains introduced by combining IRS and NOMA have been investigated in \cite{Ding,Fu,Mu_ax,Zheng}. The authors of~\cite{Ding} proposed a simple IRS-assisted NOMA transmission architecture, where the outage performance was analyzed under an on-off IRS control scheme. The authors of~\cite{Fu} investigated the joint beamforming design in downlink MISO IRS-assisted NOMA networks for minimization of the total transmit power. Moreover, the sum rate of MISO IRS-NOMA networks was maximized in~\cite{Mu_ax}, where the IRS reflection coefficients were optimized for ideal and non-ideal IRS elements. The authors of~\cite{Zheng} compared the performance of NOMA and OMA in IRS-assisted networks for different user pairing strategies. \textcolor{black}{The authors of \cite{Zhang_deployment} studied the capacity region of the two-user IRS-aided multiple access channel, which is achieved by NOMA, where both distributed and centralized IRS deployment strategies were considered.}
\subsubsection{Channel Estimation Schemes for IRS-assisted Networks}
\textcolor{black}{The optimization of the IRS reflection coefficients relies on the accurate acquisition of the channel state information (CSI) of both the direct and the reflection channels, which is challenging due to the passive nature of IRSs. The authors of~\cite{Yang_CSI} proposed a channel estimation method for IRS-aided orthogonal frequency division multiplexing (OFDM) systems based on an efficient IRS element-grouping scheme. Furthermore, the authors of~\cite{Zheng_OFDM_WCL} reported a novel IRS reflection pattern design for assisting the channel estimation in uplink IRS-aided OFDM systems. For IRS with discrete phase shifts, the authors of \cite{You_CSI} proposed a hierarchical training IRS reflection design for reflection channel estimation. In contrast to the single-user scenario considered in~\cite{Yang_CSI,Zheng_OFDM_WCL,You_CSI}, the authors of \cite{Li_CSI} investigated the channel estimation in an IRS-empowered downlink multi-user MISO system, and proposed two iterative estimation methods. Moreover, the authors of \cite{Zheng_OFDMA_WCOM} reported two channel estimation methods for IRS-assisted multi-user systems as well as corresponding channel training designs for minimizing the estimation error.}
\vspace{-0.6cm}
\subsection{Motivations and Contributions}
\vspace{-0.2cm}
One critical issue in IRS-assisted communications is that the IRS-assisted link suffers from the ``double fading'' effect~\cite{Griffin}, which causes severe path loss. \textcolor{black}{In fact, the received signal power in the IRS-assisted link in general scales with ${1 \mathord{\left/
 {\vphantom {1 {\left( {d_T^{{\alpha _T}}d_R^{{\alpha _R}}} \right)}}} \right.
 \kern-\nulldelimiterspace} {\left( {d_T^{{\alpha _T}}d_R^{{\alpha _R}}} \right)}}$~\cite{Ozdogan}, where ${{d_T}}$ and ${{d_R}}$ are the distances of the AP-IRS link and the IRS-user link, respectively, and ${{\alpha _T}}$ and ${{\alpha _R}}$ are the corresponding path loss exponents.} As a result, the performance of IRS-assisted networks is sensitive to the deployment location of the IRS. \textcolor{black}{However, there are only a few initial works~\cite{Zhang_deployment,tutorial} that have investigated the IRS deployment design. The authors of~\cite{Zhang_deployment} have recently studied optimal IRS deployment strategies for a two-user network from a \emph{network level} perspective, and distinguished centralized and distributed deployment strategies. For each strategy, the deployment locations of the IRSs are assumed to be fixed and are not optimized. In a recent tutorial~\cite{tutorial}, the optimization of the IRS deployment location was studied for an IRS-assisted single-user network, which revealed that the IRS should be deployed close to either the AP or the user. However, the solutions proposed in~\cite{Zhang_deployment,tutorial} are not applicable to the \emph{link level} optimization of the IRS deployment location in IRS-assisted multi-user networks, which is the main focus of this paper.}\\
\indent In this paper, we investigate the joint design of the IRS deployment, IRS reflection coefficients, and power allocation at the AP in a downlink IRS-assisted multi-user network for different MA schemes. The main contributions of this paper can be summarized as follows:
\begin{itemize}
  \item We investigate a downlink IRS-assisted multi-user network, in which the direct AP-user link is blocked and one IRS is deployed for coverage extension. For this system, we formulate a weighted sum rate (WSR) maximization problem for joint optimization of the deployment location and the reflection coefficients of the IRS as well as the power allocation at the AP for NOMA, frequency division multiple access (FDMA), and time division multiple access (TDMA) transmission.
  \item \textcolor{black}{For NOMA and FDMA, we develop monotonic optimization (MO) based algorithms to determine corresponding performance upper bounds,}  where the IRS reflection coefficients and the power allocation at the AP are jointly optimized using MO theory and semidefinite relaxation (SDR). Due to the \emph{time-selective} nature of the IRS, for TDMA, the optimal IRS reflection coefficients of the users can be obtained in closed form.
  \item \textcolor{black}{We further develop low-complexity alternating optimization (AO) based algorithms for all three considered MA schemes to find a high-quality suboptimal solution,} where the power allocation, IRS reflection coefficients, and deployment location are optimized in an alternating manner with the other variables fixed by employing successive convex approximation (SCA). Specifically, for IRS deployment location design, we propose a novel \emph{local region optimization} method to efficiently obtain the desired IRS deployment location. Additionally, we propose an efficient user ordering scheme for NOMA.
  \item \textcolor{black}{Our numerical results unveil that 1) the proposed suboptimal AO based algorithms achieve near-optimal performance and require much fewer iterations to converge than the MO based algorithms; 2) optimizing the IRS deployment location can significantly increase the IRS performance gain; 3) the optimal IRS deployment strategy for OMA tends to enhance the channel power gains of all users, whereas for NOMA, it is preferable that the IRS enlarges the disparities among the users' channels.}
\end{itemize}

\vspace{-0.6cm}
\subsection{Organization and Notations}
\vspace{-0.2cm}
The rest of this paper is organized as follows. Section II presents the system model and the considered WSR maximization problem formulation for downlink IRS-assisted multi-user networks. The proposed MO and AO based algorithms are developed in Section III and Section IV, respectively. Section V compares the complexity and performance of the proposed algorithms, and in Section VI, numerical results are provided to verify their effectiveness. Finally, Section VII concludes the paper. \\
\indent \emph{Notations:} Scalars are denoted by lower-case letters. Vectors and matrices are denoted by bold-face lower-case and upper-case letters, respectively. ${\mathbb{C}^{N \times 1}}$ denotes the space of $N \times 1$ complex-valued vectors. ${{\mathbf{a}}^T}$ and ${{\mathbf{a}}^H}$ denote the transpose and conjugate transpose of vector ${\mathbf{a}}$, respectively. ${\rm {diag}}\left( \mathbf{a} \right)$ is a diagonal matrix with the elements of vector ${\mathbf{a}}$ on the main diagonal. $\left\| {\mathbf{a}} \right\|$ denotes the Euclidean norm of vector ${\mathbf{a}}$. $ \otimes $ denotes the Kronecker product. ${\mathbb{H}^{N}}$ denotes the set of all $N$-dimensional complex Hermitian matrices. ${\rm {rank}}\left( \mathbf{A} \right)$ and ${\rm {Tr}}\left( \mathbf{A} \right)$ denote the rank and the trace of matrix $\mathbf{A}$, respectively. ${{\mathbf{A}}} \succeq 0$ indicates that $\mathbf{A}$ is a positive semidefinite matrix.
\vspace{-0.4cm}
\section{System Model and Problem Formulation}
\vspace{-0.2cm}
\subsection{System Model}
\vspace{-0.2cm}
\begin{figure}[h!]
    \begin{center}
        \includegraphics[width=2.8in]{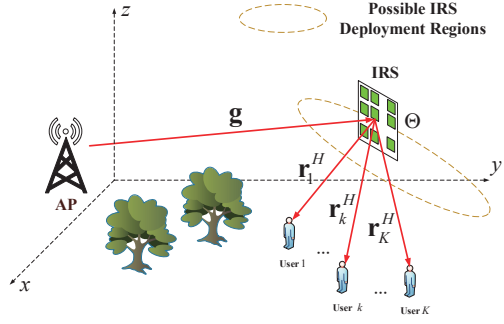}
        \caption{Illustration of the downlink IRS-assisted multi-user communication network.}
        \label{System model}
    \end{center}
\end{figure}
\vspace{-0.0cm}
\textcolor{black}{We consider a narrow-band downlink IRS-assisted communication network operating over frequency-flat channels,} and consisting of one single-antenna AP, $K$ single-antenna users, and one IRS, as illustrated in Fig. \ref{System model}. Specifically, the direct AP-user links are assumed to be blocked by obstacles\footnote{\textcolor{black}{If there is no blockage, the channel qualities of the users dependent on both the direct AP-user links and the reflection AP-IRS-user links, which results in a more challenging IRS deployment optimization problem. However, this scenario is beyond the scope of this paper and constitutes an interesting topic for future work.}}, \textcolor{black}{which is a very challenging scenario for conventional wireless communication systems.} \textcolor{black}{The IRS is deployed to provide wireless service to a given communication dead zone, where the users are assumed to be static or low-mobility, i.e., a quasi-static scenario\footnote{\textcolor{black}{If the users are widely distributed and/or high-mobility, multiple IRSs or mobile IRSs may be required to achieve high performance. The resulting deployment optimization problem is beyond the scope of the current work.}}.} Assuming a three-dimensional (3D) Cartesian coordinate system, the locations of the AP, the IRS, and the $k$th user are denoted by ${\mathbf{b}} = {\left( {{x_b},{y_b},{H_b}} \right)^T}$, ${\mathbf{s}} = {\left( {{x_s},{y_s},{H_s}} \right)^T}$, and ${{\mathbf{u}}_k} = {\left( {{x_k},{y_k},{H_k}} \right)^T}$, respectively. In this paper, let $\Omega $ specify a predefined region for deploying the IRS. The deployment location of the IRS should satisfy the following condition
\vspace{-0.3cm}
\begin{align}\label{region}
  {\mathbf{s}} \in \Omega  = \left\{ {\left( {{x_s},{y_s},{H_s}} \right)^T|{x_{\min }} \le {x_s} \le {x_{\max }},} \right. \left. {{y_{\min }} \le {y_s} \le {y_{\max }},{H_{\min }} \le {H_s} \le {H_{\max }}} \right\},
\end{align}
\vspace{-1.2cm}

\noindent where $\left[ {{x_{\min }},{x_{\max }}} \right]$, $\left[ {{y_{\min }},{y_{\max }}} \right]$, and $\left[ {{H_{\min }},{H_{\max }}} \right]$ denote the candidate ranges along the $x$-, $y$- and $z$-axes, respectively. \textcolor{black}{In practice, the size of $\Omega $ is usually limited\footnote{\textcolor{black}{Note that the solutions proposed in this paper are also applicable to large $\Omega$. This is because the IRS deployment design is an offline optimization problem, see Section II-D for details. Thus, the potentially high computational complexity caused by large $\Omega $ is acceptable given the available computing power.}}, especially for the considered quasi-static scenario, since it is not only restricted by the limited coverage of the IRS but also depends on external factors such as the line-of-sight (LoS) link requirement (e.g., the locations of surrounding obstacles) and the available infrastructure for deployment (e.g., building facades on which the IRS can be deployed).} The IRS is assumed to be equipped with a uniform planar array (UPA) composed of $M={M_v}{M_h}$ passive reflecting elements with element spacing $d_I$. \textcolor{black}{Due to the assumed narrow-band transmission, the reflection coefficients are constant across the entire signal bandwidth.} \textcolor{black}{As a result, the frequency-flat IRS reflection coefficient matrix is modelled as ${{\mathbf{\Theta}} } = {\rm {diag}}\left( {{e^{j{\theta _{1}}}},{e^{j{\theta _{2}}}}, \ldots ,{e^{j{\theta _{M}}}}} \right)$, where ${{\theta _{m}}}$ denotes the phase shift coefficient of the $m$th element of the IRS and the amplitude coefficients of all elements are set to one\footnote{\textcolor{black}{In this work, similar to~\cite{Wu2019IRS,Huang_EE}, we assume that all reflection coefficients are independent, in order to investigate the maximum achievable performance gain. However, according to a recent experimental study~\cite{Tang}, in practice, the IRS reflection coefficients are highly sensitive to the incident angle. An incident-angle-dependent reflection coefficient model makes the IRS reflection coefficient and deployment location design much more challenging. Hence, this problem is left for future work where advanced machine learning based tools could be exploited.}}.} \\
\indent In this paper, the narrow-band quasi-static fading channels of the AP-IRS link and the IRS-user links are modeled as Rician fading channels. Therefore, the small scale fading of the AP-IRS and the IRS-user links can be expressed as
\vspace{-0.3cm}
\begin{align}\label{AP-IRS}
  {\mathbf{g}} = \sqrt {\frac{\beta_{AI} }{{1 + \beta_{AI} }}} {{\mathbf{g}}^{{\rm{LoS}}}} + \sqrt {\frac{1}{{1 + \beta_{AI} }}} {{\mathbf{g}}^{{\rm{NLoS}}}} \in {{\mathbb{C}}^{M \times 1}},
\end{align}
\vspace{-0.8cm}
\begin{align}\label{IRS-User}
  {{\mathbf{r}}_k} = \sqrt {\frac{\beta_{IU} }{{1 + \beta_{IU} }}} {{\mathbf{r}}_k^{{\rm{LoS}}}} + \sqrt {\frac{1}{{1 + \beta_{IU} }}} {{\mathbf{r}}_k^{{\rm{NLoS}}}} \in {{\mathbb{C}}^{M \times 1}},
\end{align}
\vspace{-1cm}

\noindent where $\beta_{AI} $ and $\beta_{IU} $ denote the Rician factor of the AP-IRS link and IRS-user links, respectively. ${{\mathbf{g}}^{{\rm{LoS}}}} = {{\mathbf{a}}}\left( {\phi ^{AI},\varphi ^{AI}} \right)$ and ${\mathbf{r}}_k^{{\rm{LoS}}} = {{\mathbf{a}}}\left( {\phi _k^{IU},\varphi _k^{IU}} \right)$ are the deterministic line-of-sight (LoS) components, and ${{{\mathbf{g}}^{{\rm{NLoS}}}}}$ and ${{{\mathbf{r}}_k^{{\rm{NLoS}}}}}$ are the non-line-of-sight (NLoS) components modeled as Rayleigh fading. ${\mathbf{a}}\left( {\phi ,\varphi } \right)$ is the array response vector (ARV) which can be expressed as~\cite{Tse}
\vspace{-0.3cm}
\textcolor{black}{\begin{align}\label{AAR}
\begin{gathered}
  {\mathbf{a}}\left( {\phi ,\varphi } \right) = {\left[ {1, \ldots ,{e^{ - j\frac{{2\pi {d_I}}}{\lambda }\left( {{m_v} - 1} \right)\sin \phi \cos \varphi }}, \ldots ,{e^{ - j\frac{{2\pi {d_I}}}{\lambda }\left( {{M_v} - 1} \right)\sin \phi \cos \varphi }}} \right]^T} \hfill \\
   \otimes {\left[ {1, \ldots ,{e^{ - j\frac{{2\pi {d_I}}}{\lambda }\left( {{m_h} - 1} \right)\sin \phi \sin \varphi }}, \ldots ,{e^{ - j\frac{{2\pi {d_I}}}{\lambda }\left( {{M_h} - 1} \right)\sin \phi \sin \varphi }}} \right]^T}, \hfill \\
\end{gathered}
\end{align}}
\vspace{-0.8cm}

\noindent where $\phi \in \left[ { - \frac{\pi }{2},\frac{\pi }{2}} \right]$ and $\varphi  \in \left[ {0,\pi } \right]$ denote the elevation angle-of-arrival (AoA)/angle-of-departure (AoD) and the azimuth AoA/AoD, respectively. \textcolor{black}{For the elevation AoAs/AoDs, we have ${\phi ^{AI}} = \arcsin \left(\frac{{{H_b} - {H_s}}}{{\left\| {{\mathbf{b}} - {\mathbf{s}}} \right\|}}\right)$ and $\phi _k^{IU} = \arcsin \left(\frac{{{H_k} - {H_s}}}{{\left\| {{{\mathbf{u}}_k} - {\mathbf{s}}} \right\|}}\right)$, and for the azimuth AoAs/AoDs, we have ${\varphi ^{AI}} = \arccos \left( {\frac{{{x_b} - {x_s}}}{{\left\| {{{\left( {{\mathbf{b}} - {\mathbf{s}}} \right)}_{1:2}}} \right\|}}} \right)$ and $\varphi _k^{IU} = \arccos \left( {\frac{{{x_k} - {x_s}}}{{\left\| {{{\left( {{{\mathbf{u}}_k} - {\mathbf{s}}} \right)}_{1:2}}} \right\|}}} \right)$, where ${\left( \cdot    \right)_{1:2}}$ denotes the first two elements of a vector.}\\
\indent Furthermore, the path loss ${{L}_{IRS,k}}$ between the AP and user $k$ via the IRS is given by
\vspace{-0.4cm}
\begin{align}\label{product}
{L_{IRS,k}} = \frac{{\rho _0^{}}}{{d_{AI}^{{\alpha _{AI}}}}}\frac{{\rho _0^{}}}{{d_{IU,k}^{{\alpha _{IU}}}}},
\end{align}
\vspace{-1cm}

\noindent where ${{\rho _0}}$ represents the path loss at a reference distance of 1 meter, and ${d_{AI}} = \left\| {{\mathbf{s}} - {\mathbf{b}}} \right\|$ and ${d_{IU,k}} = \left\| {{\mathbf{s}} - {{\mathbf{u}}_k}} \right\|$ denote the distances of the AP-IRS link and the IRS-user $k$ link, respectively. ${{\alpha _{AI}}}$ and ${{\alpha _{IU}}}$ are the corresponding path loss exponents. From \eqref{product}, it can be observed that the IRS-assisted link suffers from the ``double-fading'' effect~\cite{Griffin}. \textcolor{black}{In this paper, we consider only signals that are reflected by the IRS one time and ignore the combinations of signals that are reflected multiple times due to the high associated path loss, see also~\cite{Wu2019IRS,Mu_ax}.} Therefore, the combined channel power gain of user $k$ can be expressed as
\vspace{-0.6cm}
\begin{align}\label{combined channel power gain}
{c_k} = {L_{IRS,k}}{\left| {{{\mathbf{r}}_k^H}{\mathbf{\Theta}} {\mathbf{g}}} \right|^2} = {\left| {{{\mathbf{q}}_k}{\mathbf{v}}} \right|^2},
\end{align}
\vspace{-1.2cm}

\noindent where ${{\mathbf{q}}_k} = \sqrt {{L_{IRS,k}}} {{\mathbf{r}}_k^H}{\rm{diag}}\left( {\mathbf{g}} \right)$ and ${\bf{v}}  = {\left[ {{e^{j{\theta _1}}},{e^{j{\theta _2}}}, \ldots ,{e^{j{\theta _M}}}} \right]^T}$.\\
\indent In this paper, we consider two different MA strategies, i.e., NOMA and OMA. For NOMA, all users share the same time and frequency resources by invoking SC at the AP and SIC at the users. For OMA, we consider TDMA and FDMA.
\vspace{-0.6cm}
\subsection{NOMA}
\vspace{-0.2cm}
Based on the NOMA principle, each user employs SIC to remove the intra-cell interference. The users with the stronger channel power gains decode first the signals of the users with the weaker channel power gains, before decoding their own signal. \textcolor{black}{Let $\mu \left( k \right) \in \left\{ {1, \ldots ,K} \right\}$ denote the decoding order of user $k$. For any two users $j$ and $k$ satisfying $\mu \left( k \right) > \mu \left( j \right)$, user $k$ will first decode the signal of user $j$ before decoding its own signal, where the combined channel power gains of the two users need to satisfy ${\left| {{\mathbf{q}}_k{\mathbf{v}}} \right|^2} \ge {\left| {{\mathbf{q}}_j{\mathbf{v}}} \right|^2}$.} Since the combined channel power gain is determined by both the IRS deployment location and the reflection coefficients, the decoding order can be any one of all $K!$ possible combinations\textcolor{black}{~\cite{Zheng,Mu_ax}}. Let ${\mathcal{D}}$ denote the set of all possible decoding orders, where $\left| {\mathcal{D}} \right| = K!$. The achievable rate of user $k$ can be expressed as
\vspace{-0.3cm}
\begin{align}\label{NOMA user k}
R_k^N = {\log _2}\left( {1 + \frac{{{{\left| {{{\mathbf{q}}_k}{\mathbf{v}}} \right|}^2}{p_k}}}{{{{\left| {{{\mathbf{q}}_k}{\mathbf{v}}} \right|}^2}\sum\nolimits_{\mu \left( i \right) > \mu \left( k \right)} {{p_i}}  + {\sigma ^2}}}} \right),
\end{align}
\vspace{-0.8cm}

\noindent where ${{p_k}}$ denotes the transmit power allocated to the $k$th user, and ${\sigma ^2}$ denotes the variance of the additive complex Gaussian noise. Let $P_{\max}$ denote the total transmit power at the AP, and we have $\sum\nolimits_{k = 1}^K {{p_k}}  \le {P_{\max}}$. \textcolor{black}{Furthermore, the allocated power should satisfy the following condition}
\vspace{-0.6cm}
\begin{align}\label{power allocation}
{p_k} \le {p_j}\;{\rm{if}}\;\mu \left( k \right) > \mu \left( j \right),
\end{align}
\vspace{-1.2cm}

\noindent \textcolor{black}{i.e., higher powers are allocated to the users with lower decoding orders and having weaker channel conditions~\cite{Liu2017}.}\\
\indent Our goal is to maximize the WSR of the users by jointly optimizing the power allocation at the AP and the reflection coefficients and deployment location of the IRS. If NOMA is employed, the optimization problem is formulated as follows
\vspace{-0.4cm}
\begin{subequations}\label{P1}
\begin{align}
({{\rm{NOMA}}}):&\mathop {\max }\limits_{\left\{ {{p_k}} \right\},{\bf{v}} ,{\mathbf{s}}} \;\;\sum\limits_{k = 1}^K {{w_k}R_k^N}    \\
\label{deployment1}{\rm{s.t.}}\;\;&{\mathbf{s}} \in \Omega ,\\
\label{IRS1}&\left| {{v_m}} \right| = 1,\forall m \in {{\mathcal{M}}},\\
\label{decoding order}&\mu \left( k \right) \in {\mathcal{D}},\forall k \in {\mathcal{K}},\\
\label{channel}&{\left| {{\mathbf{q}}_k{\mathbf{v}}} \right|^2} \ge {\left| {{\mathbf{q}}_j{\mathbf{v}}} \right|^2},{\rm{if}}\;\mu \left( k \right) > \mu \left( j \right),\\
\label{total power1}&0 \le {p_k} \le {p_j}\;{\rm{if}}\;\mu \left( k \right) > \mu \left( j \right),\\
\label{total power2}&\sum\nolimits_{k = 1}^K {{p_k}}  \le {P_{\max}},
\end{align}
\end{subequations}
\vspace{-1cm}

\noindent where ${w_k}$ is the non-negative rate weight for user $k$. Equations \eqref{deployment1} and \eqref{IRS1} are the IRS deployment location constraint and the unit modulus constraint\footnote{\textcolor{black}{In this paper, continuous phase shifts are assumed for all IRS reflection elements, which provide a performance upper bound for systems employing more practical discrete phase shifts~\cite{Huang_Holographic,Huang_Discrete}. It is worth noting that the obtained continuous phase shifts can be quantized into discrete ones and the resulting performance degradation is small for sufficiently high phase shift resolutions~\cite{Huang_Discrete,Mu_ax}.}} for the IRS reflection coefficients. Equations \eqref{decoding order} and \eqref{channel} are the NOMA decoding order constraints. Equations \eqref{total power1} and \eqref{total power2} are the constraints for the allocated powers.
\vspace{-0.6cm}
\subsection{OMA}
\vspace{-0.2cm}
\subsubsection{FDMA} For FDMA, the AP serves the users in orthogonal frequency bands of equal size \textcolor{black}{with the aid of one common frequency-flat IRS reflection coefficient vector $\mathbf{v}$.} Then, the achievable rate at user $k$ can be expressed as
\vspace{-0.3cm}
\begin{align}\label{FDMA}
R_k^F = \frac{1}{K}{\log _2}\left( {1 + \frac{{{{\left| {{{\mathbf{q}}_k}{\mathbf{v}}} \right|}^2}{p_k}}}{{\frac{1}{K}{\sigma ^2}}}} \right).
\end{align}
\vspace{-0.8cm}

\noindent Accordingly, the optimization problem for FDMA can be written as follows
\vspace{-0.3cm}
\begin{subequations}\label{P2}
\begin{align}
({{\rm{FDMA}}}):&\mathop {\max }\limits_{\left\{ {{p_k}} \right\},{\bf{v}} ,{\mathbf{s}}} \;\;\sum\limits_{k = 1}^K {{w_k}R_k^F}  \\
\label{deployment2}{\rm{s.t.}}\;\;&{\mathbf{s}} \in \Omega ,\\
\label{IRS2}&\left| {{v_m}} \right| = 1,\forall m \in {{\mathcal{M}}},\\
\label{total power3}&{p_k}\ge 0,\forall k \in {\mathcal{K}}, \sum\nolimits_{k = 1}^K {{p_k}}  \le {P_{\max}}.
\end{align}
\end{subequations}
\vspace{-1.2cm}
\subsubsection{TDMA} For TDMA, the AP serves the users in orthogonal time slots of equal size. We note that the IRS reflection coefficients can assume different values in each time slot. \textcolor{black}{This property of IRSs is referred to as \emph{time-selectivity}~\cite{Zheng}. Because of this time-selectivity, TDMA is different from NOMA and FDMA, where the IRS reflection coefficients are always identical for all users}\footnote{\textcolor{black}{We note that TDMA entails a higher hardware complexity for the IRS than NOMA/FDMA, since the IRS reflection coefficients need to be reconfigured multiple times during the transmission. This, however, is a non-trivial task since the size of the IRS is usually large, which makes the IRS reconfiguration time-consuming.}}$^,$\footnote{\textcolor{black}{In fact, in principle, NOMA and FDMA can also exploit the \emph{time-selectivity} of the IRS. To facilitate this, the IRS reflection coefficients and the corresponding power allocation have to be jointly optimized in each time slot. However, these considerations are beyond the scope of this work.}}. As a result, for TDMA, the achievable rate at user $k$ can be expressed as
\vspace{-0.3cm}
\begin{align}\label{TDMA}
R_k^T = \frac{1}{K}{\log _2}\left( {1 + \frac{{{{\left| {{{\mathbf{q}}_k}{{\mathbf{v}}_k}} \right|}^2}{P_{\max}}}}{{{\sigma ^2}}}} \right),
\end{align}
\vspace{-0.8cm}

\noindent where ${{\bf{v}} _k}$ represents the IRS reflection coefficients for user $k$. As the AP transmits to only one user at a given time, the transmit power is always set as ${P_{\max}}$.\\
\indent Then, the optimization problem for TDMA can be formulated as
\vspace{-0.3cm}
\begin{subequations}\label{P3}
\begin{align}
({{\rm{TDMA}}}):&\mathop {\max }\limits_{\left\{ {{{\bf{v}}_k}} \right\} ,{\mathbf{s}}} \;\;\sum\limits_{k = 1}^K {{w_k}R_k^T}    \\
\label{deployment}{\rm{s.t.}}\;\;&{\mathbf{s}} \in \Omega ,\\
\label{IRS3}&\left| {{v_{k,m}}} \right| = 1,\forall k \in {{\mathcal{K}}},\forall m \in {{\mathcal{M}}}.
\end{align}
\end{subequations}
\vspace{-1.8cm}
\subsection{Discussion}
\vspace{-0.2cm}
\textcolor{black}{Note that since the random NLoS components ${{{\mathbf{g}}^{{\rm{NLoS}}}}}$ and ${{{\mathbf{r}}_k^{{\rm{NLoS}}}}}$ of the IRS-assisted links are impossible to obtain before the deployment of the IRS, the formulated joint optimization problems \eqref{P1}, \eqref{P2}, and \eqref{P3} can be solved only for deterministic channel coefficients. Since the IRS is usually deployed to avoid signal blockage, the deterministic LoS components are expected to be the dominant factor. Motivated by this, we first solve the proposed joint optimization problems \textcolor{black}{in an offline manner based on the LoS components to determine a favorable IRS deployment location.} After deployment of the IRS, the IRS reflection coefficients and the power allocation can be obtained by solving problems \eqref{P1}, \eqref{P2}, and \eqref{P3} again \textcolor{black}{in an online manner for the given deployment location and instantaneous CSI (I-CSI).} \textcolor{black}{I-CSI can be efficiently obtained with one of the recently proposed channel estimation methods for IRS-assisted multi-user networks, e.g.,~\cite{Li_CSI,Zheng_OFDMA_WCOM}.} In this paper, we mainly focus on the first problem since the second optimization problem for the deployed IRS and I-CSI can be solved in a similar manner. Note that the WSRs obtained for both problems will be similar, when the Rician factor is large. In the following, we develop MO and AO based algorithms to find performance upper bounds and high-quality suboptimal solutions, respectively.}
%
%
\vspace{-0.6cm}
\section{Monotonic Optimization Based Algorithms}
\vspace{-0.2cm}
In this section, we develop MO based algorithms which find a performance upper bound for the formulated NOMA and FDMA optimization problems. Specifically, we transform these non-convex optimization problems into the canonical form of MO problems, which are solved by invoking the polyblock outer approximation algorithm and SDR. Then, we optimally solve the TDMA optimization problem in closed form by exploring the time-selective nature of the IRS.
\vspace{-0.6cm}
\subsection{NOMA}
\vspace{-0.2cm}
The optimal WSR for NOMA can be obtained by exhaustively searching over all possible IRS locations $\Omega $ and decoding orders ${{\mathcal{D}}}$, i.e., ${\mathcal{R}}_N^* = \mathop {\max }\limits_{{\mathbf{s}} \in \Omega ,\left\{ {{\mu _k}} \right\} \in {\mathcal{D}}} \sum\limits_{k = 1}^K {{w_k}R_k^{N*}\left( {{\mathbf{s}},\left\{ {{\mu _k}} \right\}} \right)} $, where ${R_k^{N*}\left( {{\mathbf{s}},\left\{ {{\mu _k}} \right\}} \right)}$ denotes the optimal solution for a given IRS deployment location and user decoding order. In the following, we solve the corresponding subproblem by employing monotonic optimization.
\subsubsection{\textbf{Monotonic Optimization Problem Transformation}} Define ${{\mathbf{Q}}_k} = {\mathbf{q}}_k^H{\mathbf{q}}_k^{},\forall k \in {\mathcal{K}}$, and ${\mathbf{V}} = {\mathbf{v}}{{\mathbf{v}}^H}$ which satisfies ${{\mathbf{V}}} \succeq 0$, ${\rm {rank}}\left( {{{\mathbf{V}}}} \right) = 1$ and ${\left[ {\mathbf{V}} \right]_{mm}} = 1,m = 1,2, \ldots ,M$. The achievable rate at user $k$ in \eqref{NOMA user k} can be rewritten as follows
\vspace{-0.3cm}
\begin{align}\label{NOMA user k SDR}
  R_k^N = {\log _2}\left( {1 + \frac{{{\rm{Tr}}\left( {{\mathbf{V}}{{\mathbf{Q}}_k}} \right){p_k}}}{{\sum\nolimits_{\mu \left( i \right) > \mu \left( k \right)} {{\rm{Tr}}\left( {{\mathbf{V}}{{\mathbf{Q}}_k}} \right){p_i}}  + {\sigma ^2}}}} \right) = {\log _2}\left( {1 + \frac{{{p_k}}}{{\sum\nolimits_{\mu \left( i \right) > \mu \left( k \right)} {{p_i}}  + \frac{{{\sigma ^2}}}{{{\rm{Tr}}\left( {{\mathbf{V}}{{\mathbf{Q}}_k}} \right)}}}}} \right).
\end{align}
\vspace{-0.8cm}

\noindent Then, we introduce auxiliary variables ${\lambda _k}$ such that ${\lambda _k} = \frac{{{\sigma ^2}}}{{{\rm{Tr}}\left( {{\mathbf{V}}{{\mathbf{Q}}_k}} \right)}},\forall k \in {\mathcal{K}}$. For a given IRS deployment location and user decoding order, the resulting subproblem of \eqref{P1} can be expressed as
\vspace{-0.3cm}
\begin{subequations}\label{P1-order-location}
\begin{align}
({{\rm{N-sub}}}):&\mathop {\max }\limits_{\left\{ {{p_k},{\lambda _k}} \right\},{\mathbf{V}}} \;\;\sum\limits_{k = 1}^K {{w_k}{{\log }_2}} \left( {1 + \frac{{{p_k}}}{{\sum\nolimits_{\mu \left( i \right) > \mu \left( k \right)} {{p_i}}  + {\lambda _k}}}} \right)    \\
\label{total power1-order}{\rm{s.t.}}\;\;&0 \le {p_k} \le {p_j}\;{\rm{if}}\;\mu \left( k \right) > \mu \left( j \right),\\
\label{total power2-order}&\sum\nolimits_{k = 1}^K {{p_k}}  \le {P_{\max}},\\
\label{lamda}&{\rm{Tr}}\left( {{\mathbf{V}}{{\mathbf{Q}}_k}} \right) \ge \frac{{{\sigma ^2}}}{{{\lambda _k}}},\forall k \in {\mathcal{K}},\\
\label{channel-order}&{{\rm{Tr}}\left( {{\mathbf{V}}{{\mathbf{Q}}_k}} \right)} \ge {{\rm{Tr}}\left( {{\mathbf{V}}{{\mathbf{Q}}_j}} \right)},{\rm{if}}\;\mu \left( k \right) > \mu \left( j \right),\\
\label{Vmm}&{\left[ {\mathbf{V}} \right]_{mm}} = 1,\forall m \in {{\mathcal{M}}},{{\mathbf{V}}}  \succeq  0, {\mathbf{V}} \in {{\mathbb{H}}^{M}},\\
\label{rank 1 V}&{\rm {rank}}\left( {{{\mathbf{V}}}} \right) = 1.
\end{align}
\end{subequations}
\vspace{-1.2cm}

\noindent Note that the inequality constraint \eqref{lamda} does not affect the equivalence of problem \eqref{P1-order-location} and the original problem \eqref{P1}. At optimality, constraint \eqref{lamda} is always met with equality. To demonstrate this, assume that one of the constraints in \eqref{lamda} is satisfied with strict inequality. Then, we can always decrease that ${{\lambda _k}}$ to satisfy the constraint with equality, which also increases the objective value. Therefore, for the optimal solution of problem \eqref{P1-order-location}, constraint \eqref{lamda} must be satisfied with equality.\\
\indent To facilitate the application of MO, we introduce auxiliary variables ${\gamma _k}$ which satisfy the following constraint
\vspace{-0.6cm}
\begin{align}\label{auxiliary variables}
1 \le {\gamma _k} \le 1 + \frac{{{p_k}}}{{\sum\nolimits_{\mu \left( i \right) > \mu \left( k \right)} {{p_i}}  + {\lambda _k}}},\forall k \in {\mathcal{K}}.
\end{align}
\vspace{-1cm}

\noindent Then, problem \eqref{P1-order-location} can be equivalently rewritten as
\vspace{-0.3cm}
\begin{subequations}\label{P1-order-location mo}
\begin{align}
&\mathop {\max }\limits_{\bm{\gamma}}  \;\;\sum\limits_{k = 1}^K {{w_k}{{\log }_2}} \left( {{\gamma _k}} \right) \\
\label{con1}{\rm{s.t.}}\;\;&{\bm{\gamma}}  \in {\mathcal{G}} \cap {\mathcal{H}},
\end{align}
\end{subequations}
\vspace{-1cm}

\noindent where ${\bm{\gamma}}  = {\left[ {{\gamma _1},{\gamma _2}, \ldots ,{\gamma _K}} \right]^T} \in {{\mathbb{R}}^{K \times 1}}$ and ${\mathcal{G}} \cap {\mathcal{H}}$ denotes the feasible set. Specifically, ${\mathcal{G}}$ and ${\mathcal{H}}$ are a normal set and a conormal set, respectively~\cite{MO}, which are given by
\vspace{-0.3cm}
\begin{align}\label{G}
{\mathcal{G}} = \left\{ {{\bm{\gamma}} |0 \le {\gamma _k} \le 1 + \frac{{{p_k}}}{{\sum\nolimits_{\mu \left( i \right) > \mu \left( k \right)} {{p_i}}  + {\lambda _k}}},\forall k \in {\mathcal{K}},\left( {\left\{ {{p_k},{\lambda _k}} \right\},{\mathbf{V}}} \right) \in {\mathcal{F}}} \right\},
\end{align}
\vspace{-1.2cm}
\begin{align}\label{H}
{\mathcal{H}} = \left\{ {{\bm{\gamma}} |1 \le {\gamma _k},\forall k \in {\mathcal{K}}} \right\},
\end{align}
\vspace{-1.2cm}

\noindent where feasible set ${\mathcal{F}}$ is spanned by constraints \eqref{total power1-order}-\eqref{rank 1 V}. \textcolor{black}{The equivalence of problems \eqref{P1-order-location} and \eqref{P1-order-location mo} can be shown as follows. Firstly, the definition of ${\mathcal{F}}$ ensures that the optimization variables $\left( {\left\{ {{p_k},{\lambda _k}} \right\},{\mathbf{V}}} \right)$ in both problems \eqref{P1-order-location} and \eqref{P1-order-location mo} belong to the same feasible set. Furthermore, it can be verified that the condition ${\gamma _k} = 1 + \frac{{{p_k}}}{{\sum\nolimits_{\mu \left( i \right) > \mu \left( k \right)} {{p_i}}  + {\lambda _k}}}$ is satisfied for each $k\in {\mathcal{K}}$ at the optimal solution of \eqref{P1-order-location mo}. Otherwise, the objective function's value in \eqref{P1-order-location mo} would decrease, which contradicts the optimality. Therefore, the introduction of auxiliary variables ${\gamma _k}$ does not affect the equivalence of problems \eqref{P1-order-location} and \eqref{P1-order-location mo}.}
\subsubsection{\textbf{Polyblock Outer Approximation Algorithm}} \textcolor{black}{With the above equivalent transformation, problem \eqref{P1-order-location mo} is in the canonical form of a monotonic optimization problem~\cite{MO}.} Therefore, the optimal solution of problem \eqref{P1-order-location mo} is on the upper boundary of feasible set ${\mathcal{G}} \cap {\mathcal{H}}$, which can be obtained by invoking the polyblock outer approximation algorithm~\cite{MO}. To facilitate this algorithm, we first initialize a polyblock ${{{\mathcal{P}}^{\left( 1 \right)}}}$ that contains the feasible set ${\mathcal{G}} \cap {\mathcal{H}}$. The vertex of the polyblock ${{{\mathcal{P}}^{\left( 1 \right)}}}$ is defined as ${{\mathbf{z}}^{\left( 1 \right)}} = {\left[ {\gamma _1^{\left( 1 \right)},\gamma _2^{\left( 1 \right)}, \ldots ,\gamma _K^{\left( 1 \right)}} \right]^T}$. Let ${{\mathcal{T}}^{\left( 1 \right)}} = \left\{ {{{\mathbf{z}}^{\left( 1 \right)}}} \right\}$ denote the vertex set of polyblock ${{{\mathcal{P}}^{\left( 1 \right)}}}$. Based on vertex ${{{\mathbf{z}}^{\left( 1 \right)}}}$, we can generate $K$ new vertices as follows:
\vspace{-0.4cm}
\begin{align}\label{vertex generation}
\widetilde {\mathbf{z}}_k^{\left( 1 \right)} = {{\mathbf{z}}^{\left( 1 \right)}} - \left( {z_k^{\left( 1 \right)} - {\pi _k}\left( {{{\mathbf{z}}^{\left( 1 \right)}}} \right)} \right){{\mathbf{e}}_k},k = 1, \ldots ,K,
\end{align}
\vspace{-1.2cm}

\noindent where ${z_k^{\left( 1 \right)}}$ and ${{\pi _k}\left( {{{\mathbf{z}}^{\left( 1 \right)}}} \right)}$ denote the $k$th elements of vectors ${{\mathbf{z}}^{\left( 1 \right)}}$ and ${{\pi}\left( {{{\mathbf{z}}^{\left( 1 \right)}}} \right)}$, respectively. ${{\pi}\left( {{{\mathbf{z}}^{\left( 1 \right)}}} \right)}$ is the projection of ${{{\mathbf{z}}^{\left( 1 \right)}}}$ onto set ${\mathcal{G}}$ and ${{\mathbf{e}}_k}$ denotes the unit vector whose $k$th element is 1. Then, the new vertex set is given by
\vspace{-0.4cm}
\begin{align}\label{new vertex set}
{{\mathcal{T}}^{\left( 2 \right)}} = {{\mathcal{T}}^{\left( 1 \right)}}\backslash {{\mathbf{z}}^{\left( 1 \right)}} \cup \left\{ {\widetilde {\mathbf{z}}_k^{\left( 1 \right)},k = 1, \ldots ,K} \right\}.
\end{align}
\vspace{-1.2cm}

\noindent Let ${{{\mathcal{P}}^{\left( 2 \right)}}}$ denote the new polyblock defined by vertex set ${{\mathcal{T}}^{\left( 2 \right)}}$. By doing so, polyblock ${{{\mathcal{P}}^{\left( 1 \right)}}}$ is reduced to ${{{\mathcal{P}}^{\left( 2 \right)}}}$, which still contains the feasible set ${\mathcal{G}} \cap {\mathcal{H}}$. For the new polyblock ${{{\mathcal{P}}^{\left( 2 \right)}}}$, the vertex that achieves the maximum objective value is selected as the optimal vertex of ${{{\mathcal{P}}^{\left( 2 \right)}}}$ as follows:
\vspace{-0.6cm}
\begin{align}\label{new vertex}
{{\mathbf{z}}^{\left( 2 \right)}} = \mathop {\arg \max }\limits_{{\mathbf{z}} \in {{\mathcal{T}}^{\left( 2 \right)}} \cap {\mathcal{H}}} \;\sum\limits_{k = 1}^K {w_k}{{{\log }_2}\left( {{\gamma _k}} \right)}.
\end{align}
\vspace{-1cm}

\noindent By repeating the above procedure, we can successively construct a sequence of polyblocks such that
\vspace{-0.7cm}
\begin{align}\label{shrink}
{{\mathcal{P}}^{\left( 1 \right)}} \supset {{\mathcal{P}}^{\left( 2 \right)}} \supset  \ldots  \supset {\mathcal{G}} \cap {\mathcal{H}}.
\end{align}
\vspace{-1.2cm}

\begin{figure}[t!]
    \begin{center}
        \includegraphics[width=5 in,height=1.5 in]{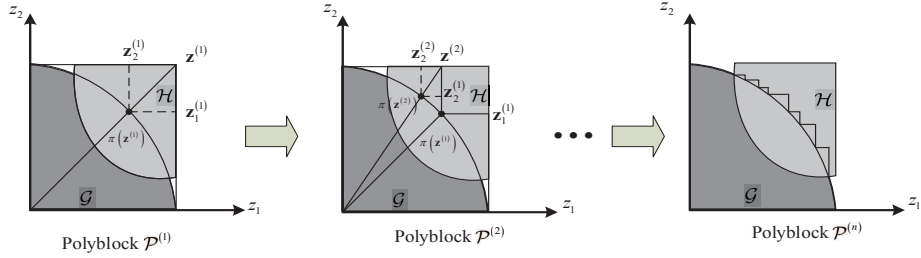}
        \caption{\textcolor{black}{Illustration of the polyblock outer approximation algorithm for the two-user case.}}
        \label{POA}
    \end{center}
\end{figure}

\noindent We illustrate the generation of the polyblocks for the two user case in Fig. \ref{POA}. Let $U\left( {{{\mathbf{z}}^{\left( n \right)}}} \right)$ and $U\left( {{{\bm{\gamma}} ^*}} \right)$ denote the objective value achieved by the vertex ${{{\mathbf{z}}^{\left( n \right)}}}$ and the best feasible solution in the $n$th iteration. The algorithm terminates when $\left| {U\left( {{{\mathbf{z}}^{\left( n \right)}}} \right) - U\left( {{{\bm{\gamma}} ^*}} \right)} \right| \le \varepsilon $, where $\varepsilon>0$ is a given tolerance. The details of the polyblock outer approximation algorithm are summarized in \textbf{Algorithm 1}.\\
\begin{algorithm}[t!]\label{method1}
\caption{Polyblock Outer Approximation Algorithm for Solving Problem \eqref{P1-order-location mo}}
 \hspace*{0.02in}
\hspace*{0.02in}{Initialize polyblock ${{{\mathcal{P}}^{\left( 1 \right)}}}$ and vertex ${{\mathbf{z}}^{\left( 1 \right)}} = {\left[ {\gamma _1^{\left( 1 \right)},\gamma _2^{\left( 1 \right)}, \ldots ,\gamma _K^{\left( 1 \right)}} \right]^T}$, where $\gamma _k^{\left( 1 \right)} = 1 + \frac{{{P_{\max }}\left\| {{{\mathbf{q}}_k}} \right\|_1^2}}{{{\sigma ^2}}},\forall k \in {\mathcal{K}}$. ${{\mathcal{T}}^{\left( 1 \right)}} = \left\{ {{{\mathbf{z}}^{\left( 1 \right)}}} \right\}$. \\ Define tolerance $\varepsilon $, $U\left( {{{\bm{\gamma}} ^*}} \right) = 0$, and $n=0$.}\\
\vspace{-0.4cm}
\begin{algorithmic}[1]
\STATE {\bf repeat}
\STATE $n=n+1$.
\STATE ${{\mathbf{z}}^{\left( n \right)}} = \mathop {\arg \max }\limits_{{\mathbf{z}} \in {{\mathcal{T}}^{\left( n \right)}} \cap {\mathcal{H}}} \;\sum\limits_{k = 1}^K {{{\log }_2}\left( {{\gamma _k}} \right)} $.
\STATE Find $\pi \left( {{{\mathbf{z}}^{\left( n \right)}}} \right)$, the projection of vertex ${{\mathbf{z}}^{\left( n \right)}}$ with \textbf{Algorithm 2}.
\STATE {\bf if} $U\left( {\pi \left( {{{\mathbf{z}}^{\left( n \right)}}} \right)} \right) \ge U\left( {{{\bm{\gamma}} ^*}} \right)$ and $\pi \left( {{{\mathbf{z}}^{\left( n \right)}}} \right) \in {\mathcal{G}} \cap {\mathcal{H}}$ {\bf then}
\STATE ${{\bm{\gamma}} ^*} = {{\mathbf{z}}^{\left( n \right)}}$ and $U\left( {{{\bm{\gamma}}  ^*}} \right) = U\left( {\pi \left( {{{\mathbf{z}}^{\left( n \right)}}} \right)} \right)$.
\STATE {\bf end if}
\STATE Generate $K$ new vertices with $\pi \left( {{{\mathbf{z}}^{\left( n \right)}}} \right)$ as $\widetilde {\mathbf{z}}_k^{\left( n \right)} = {{\mathbf{z}}^{\left( n \right)}} - \left( {z_k^{\left( n \right)} - {\pi _k}\left( {{{\mathbf{z}}^{\left( n \right)}}} \right)} \right){{\mathbf{e}}_k},k = 1, \ldots ,K$.
\STATE Construct a smaller polyblock ${{{\mathcal{P}}^{\left( n+1 \right)}}}$ with the vertex set ${{\mathcal{T}}^{\left( {n + 1} \right)}} = {{\mathcal{T}}^{\left( n \right)}}\backslash {{\mathbf{z}}^{\left( n \right)}} \cup \left\{ {\widetilde {\mathbf{z}}_k^{\left( n \right)},k = 1, \ldots ,K} \right\}$.
\STATE {\bf until} $\left| {U\left( {{{\mathbf{z}}^{\left( n \right)}}} \right) - U\left( {{{\bm{\gamma}} ^*}} \right)} \right| \le \varepsilon $
\end{algorithmic}
\hspace*{0.02in}{Obtain the optimal solution $\left\{ {p_k^*,\lambda _k^*} \right\}$ and ${{{\mathbf{V}}^*}}$ with ${{\bm{\gamma}} ^*}$. }
\end{algorithm}
\indent \textcolor{black}{Next, we explain how to initialize vertex ${{{\mathbf{z}}^{\left( 1 \right)}}}$ of polyblock ${{{\mathcal{P}}^{\left( 1 \right)}}}$. Since ${{{\mathcal{P}}^{\left( 1 \right)}}}$ should contain the feasible set ${\mathcal{G}} \cap {\mathcal{H}}$, for any user $k$, $\gamma _k^{\left( 1 \right)}$ can be set to its maximum value, which can be obtained by 1) allocating all transmit power $P_{\max}$ to user $k$; and 2) maximizing the combined channel power gain of user $k$, ${\left| {{{\mathbf{q}}_k}{\mathbf{v}}} \right|^2}$, using the corresponding optimal IRS reflection coefficients. Then, we have}
\vspace{-0.4cm}
\textcolor{black}{\begin{align}\label{in}
\gamma _k^{\left( 1 \right)} = \mathop {\max }\limits_{{{\mathbf{v}}}} \left( {1 + \frac{{{P_{\max }}{{\left| {{{\mathbf{q}}_k}{{\mathbf{v}}}} \right|}^2}}}{{{\sigma ^2}}}} \right)\mathop  = \limits^{\left( a \right)} 1 + \frac{{{P_{\max }}\left\| {{{\mathbf{q}}_k}} \right\|_1^2}}{{{\sigma ^2}}},\forall k \in {\mathcal{K}}.
\end{align}}
\vspace{-0.8cm}

\noindent \textcolor{black}{Here, ${\left( a \right)}$ holds due to the fact that ${\left| {{{\mathbf{q}}_k}{\mathbf{v}}} \right|^2} \le {\left( {\sum\nolimits_{m = 1}^M {\left| {{{\left[ {{{\mathbf{q}}_k}} \right]}_m}} \right|} } \right)^2} = \left\| {{{\mathbf{q}}_k}} \right\|_1^2$, where the inequality holds with equality when the IRS reflection coefficients ${\mathbf{v}}$ are chosen as follows:}
%
\vspace{-0.4cm}
\textcolor{black}{\begin{align}\label{ANGEL}
v_{m}^* = {e^{j\left( {{\theta ^*} - {\rm{angle}}\left( {{{\left[ {{\mathbf{q}}_k} \right]}_m}} \right)} \right)}},\forall m \in {\mathcal{M}},
\end{align}}
\vspace{-1.2cm}

\noindent \textcolor{black}{where ${{\rm{angle}}\left( {{{\left[ {{\mathbf{q}}_k} \right]}_m}} \right)}$ denotes the phase of the $m$th element of ${{\mathbf{q}}_k}$. ${{\theta ^*}}$ can be set to any arbitrary value, and we set ${\theta ^*} = 0$ for simplicity. As a result, the initial vertex ${{{\mathbf{z}}^{\left( 1 \right)}}}$ is obtained with \eqref{in} and \eqref{ANGEL} to facilitate the polyblock outer approximation algorithm.}
\subsubsection{\textbf{Finding the projection of vertex ${{\mathbf{z}}^{\left( n \right)}}$}} In each iteration of \textbf{Algorithm 1}, we need to find the projection of vertex ${{\mathbf{z}}^{\left( n \right)}}$ onto the set ${\mathcal{G}}$, which is given by $\pi \left( {{{\mathbf{z}}^{\left( n \right)}}} \right) = \alpha {{\mathbf{z}}^{\left( n \right)}},\alpha  \in \left[ {0,1} \right]$. Then, the projection can be obtained by solving the following problem
\vspace{-0.4cm}
\begin{align}\label{a}
{\alpha ^*} = \max \left\{ {{\overline \alpha  } |{\overline \alpha  } {{\mathbf{z}}^{\left( n \right)}} \in {\mathcal{G}}} \right\}.
\end{align}
\vspace{-1.2cm}

\noindent Therefore, the optimal ${\alpha ^*}$ can be obtained by invoking the bisection search method. For a given projection parameter ${\overline \alpha  }$ and vertex ${{\mathbf{z}}^{\left( n \right)}}$, the feasibility of ${\overline \alpha  {{\mathbf{z}}^{\left( n \right)}} \in {\mathcal{G}}}$ can be checked by solving the following convex problem
\vspace{-0.7cm}
\begin{subequations}\label{check}
\begin{align}
&\mathop {\max }\limits_{\left\{ {{p_k},{\lambda _k}} \right\},{\mathbf{V}}} \;\;1    \\
\label{a1}{\rm{s.t.}}\;\;&\left( {\overline \alpha  {\mathbf{z}}_k^{\left( n \right)} - 1} \right)\left( {\sum\nolimits_{\mu \left( i \right) > \mu \left( k \right)} {{p_i}}  + {\lambda _k}} \right) \le {p_k},\forall k \in {\mathcal{K}},\\
\label{cona}&\eqref{total power1-order}-\eqref{rank 1 V}.
\end{align}
\end{subequations}
\vspace{-1.4cm}

\noindent Note that problem \eqref{check} is non-convex due to the rank-one constraint \eqref{rank 1 V}. To tackle this issue, we apply SDR by relaxing this constraint. Then, problem \eqref{check} is a convex problem, which can be efficiently solved by standard convex optimization solvers such as CVX~\cite{cvx}. The procedure to find the projection of vertex ${{\mathbf{z}}^{\left( n \right)}}$ is summarized in \textbf{Algorithm 2}.
\vspace{-0.5cm}
\begin{algorithm}[!h]\label{method2}
\caption{Bisection Search Algorithm to Find the Projection of Vertex ${{\mathbf{z}}^{\left( n \right)}}$}
 \hspace*{0.02in}
\hspace*{0.02in} {Initialize $\alpha_{\max}=1$, $\alpha_{\min}=0$, and tolerance $\varepsilon $.}\\
\vspace{-0.4cm}
\begin{algorithmic}[1]
\STATE {\bf while} ${\alpha_{\max }} - {\alpha_{\min }} \ge \varepsilon $ {\bf do}
\STATE ~~Check the feasibility of problem \eqref{check} for given ${\overline \alpha  } = \frac{{{\alpha_{\max }} + {\alpha_{\min }}}}{2}$ and ${{\mathbf{z}}^{\left( n \right)}}$.
\STATE ~~{\bf if} problem \eqref{check} is feasible, {\bf then}
\STATE ~~~~${\alpha_{\min }} = {\overline \alpha  }$.
\STATE ~~{\bf else}
\STATE ~~~~${\alpha_{\max }} = {\overline \alpha  }$.
\STATE ~~{\bf end if}
\STATE {\bf end while}
\STATE ${\alpha ^*} = {\alpha _{\min }}$ and the projection is $\pi \left( {{{\mathbf{z}}^{\left( n \right)}}} \right) = {\alpha ^*}{{\mathbf{z}}^{\left( n \right)}}$. The corresponding optimization variables $\left\{ {p_k,\lambda _k} \right\}$ and ${{{\mathbf{V}}}}$ are obtained by solving problem \eqref{check} with ${\alpha ^*}$.
\end{algorithmic}
\end{algorithm}
\vspace{-1cm}
\textcolor{black}{\begin{remark}\label{SDR1}
\emph{The solution obtained with \textbf{Algorithm 1} corresponds in general to an upper bound for the optimal solution of the original problem \eqref{P1-order-location}. This is because the relaxation of the non-convex rank constraint enlarges the feasible set of \eqref{P1-order-location}. This upper bound provides a performance benchmark for validation of the optimality of any suboptimal solution.}
\end{remark}}
\vspace{-0.8cm}
\begin{remark}\label{SDR2}
\emph{When the solution ${{\mathbf{V}}^*}$ obtained with \textbf{Algorithm 1} is not rank-one, we can construct a rank-one solution using the Gaussian randomization method~\cite{Luo}. Then, the reflection coefficients $ {{{\mathbf{v}}}} $ can be obtained through Cholesky decomposition.}
\end{remark}
\vspace{-0.8cm}
\begin{remark}\label{SDR3}
\emph{\textcolor{black}{After deploying the IRS at the obtained location ${{\mathbf{s}}^*}$, the IRS reflection coefficients and power allocation can be optimized again with \textbf{Algorithm 1} based on I-CSI to find a performance upper bound.}}
\end{remark}
\vspace{-0.8cm}
\subsection{OMA}
\vspace{-0.2cm}
The optimal WSR for FDMA and TDMA can also be obtained by exhaustively searching over all possible IRS deployment locations. In this subsection, we focus again on the subproblem resulting for a given IRS deployment location.
\subsubsection{\textbf{FDMA}} With the auxiliary variables ${\left\{ {{\lambda _k^F}} \right\}}$, the subproblem for FDMA for a given IRS deployment location is equivalent to the following problem
\vspace{-0.4cm}
\begin{subequations}\label{P2-order-location}
\begin{align}
({{\rm{F-sub}}}):&\mathop {\max }\limits_{\left\{ {{p_k},{\lambda _k^F}} \right\},{\mathbf{V}}} \;\;\sum\limits_{k = 1}^K {\frac{{{w_k}}}{K}{{\log }_2}} \left( {1 + \frac{{{p_k}}}{{{\lambda _k^F}}}} \right)    \\
\label{total powerF1}{\rm{s.t.}}\;\;&{p_k}\ge 0,\forall k \in {\mathcal{K}}, \sum\nolimits_{k = 1}^K {{p_k}}  \le {P_{\max}},\\
\label{lamdaF}&{\rm{Tr}}\left( {{\mathbf{V}}{{\mathbf{Q}}_k}} \right) \ge \frac{{{\sigma ^2}}}{{K{\lambda _k^F}}},\forall k \in {\mathcal{K}},\\
\label{VmmF}&{\left[ {\mathbf{V}} \right]_{mm}} = 1,\forall m \in {{\mathcal{M}}},{{\mathbf{V}}}  \succeq  0, {\mathbf{V}} \in {{\mathbb{H}}^{M}},\\
\label{rank 1 VF}&{\rm {rank}}\left( {{{\mathbf{V}}}} \right) = 1.
\end{align}
\end{subequations}
\vspace{-1.2cm}

\noindent Note that problem \eqref{P2-order-location} has a similar structure as problem \eqref{P1-order-location}. Thus, we can also invoke MO theory to solve it. With the auxiliary variables $\left\{ {\gamma _k^F} \right\}$, problem \eqref{P2-order-location} can be rewritten as follows:
\vspace{-0.4cm}
\begin{subequations}\label{P2-order-location mo}
\begin{align}
&\mathop {\max }\limits_{\bm{\gamma}^{F}}  \;\;\sum\limits_{k = 1}^K {{\frac{{{w_k}}}{K}}{{\log }_2}} \left( {{\gamma _k^F}} \right) \\
\label{con11}{\rm{s.t.}}\;\;&{\bm{\gamma}^{F}}  \in {\mathcal{G}^{F}} \cap {\mathcal{H}^{F}},
\end{align}
\end{subequations}
\vspace{-1.2cm}

\noindent where normal set ${\mathcal{G}^{F}}$ and conormal set ${\mathcal{H}^{F}}$ are given by
\vspace{-0.4cm}
\begin{align}\label{Gf}
{\mathcal{G}^{F}} = \left\{ {{\bm{\gamma}} |0 \le {\gamma _k^F} \le 1 + \frac{{{p_k}}}{{ {\lambda _k^F}}},\forall k \in {\mathcal{K}},\left( {\left\{ {{p_k},{\lambda _k^F}} \right\},{\mathbf{V}}} \right) \in {\mathcal{U}}} \right\},
\end{align}
\vspace{-1.4cm}
\begin{align}\label{Hf}
{\mathcal{H}^{F}} = \left\{ {{\bm{\gamma}^F} |1 \le {\gamma _k^F},\forall k \in {\mathcal{K}}} \right\}.
\end{align}
\vspace{-1.2cm}

\noindent Here, feasible set ${\mathcal{U}}$ is spanned by constraints \eqref{total powerF1}-\eqref{rank 1 VF}. Problem \eqref{P2-order-location mo} can be solved again with the polyblock outer approximation algorithm. The details are omitted here for brevity.
\subsubsection{\textbf{TDMA}} As explained before, for TDMA, the AP can transmit to all users with unique IRS reflection coefficients in different time slots. For a given IRS deployment location, the WSR subproblem for TDMA can be formulated as
\vspace{-0.4cm}
\begin{subequations}\label{P3 sub}
\begin{align}
({{\rm{T-sub}}}):&\mathop {\max }\limits_{\left\{ {{{\mathbf{v}}_k}} \right\}} \;\sum\limits_{k = 1}^K {\frac{{{w_k}}}{K}{{\log }_2}\left( {1 + \frac{{{P_{\max }}{{\left| {{\mathbf{q}}_k{{\mathbf{v}}_k}} \right|}^2}}}{{{\sigma ^2}}}} \right)}      \\
\label{IRS3 sub}{\rm{s.t.}}\;\;&\left| {{v_{k,m}}} \right| = 1,\forall k \in {{\mathcal{K}}},\forall m \in {{\mathcal{M}}}.
\end{align}
\end{subequations}
\vspace{-1.2cm}

\noindent Problem \eqref{P3 sub} can be further decomposed into $K$ independent subproblems. The optimal solution to each subproblem is given by \eqref{ANGEL}. As a result, problem \eqref{P3 sub} can be solved globally optimally in closed form.
%
\vspace{-0.6cm}
\section{Alternating Optimization Based Algorithms}
\vspace{-0.2cm}
Though the proposed MO based algorithms provide performance upper bounds, the computational complexity of the polyblock outer approximation increases exponentially with the number of users. Additionally, the exhaustive search over all possible NOMA decoding orders and IRS deployment locations also entails a prohibitive complexity, which is unaffordable in practical applications. In this section, we propose low-complexity suboptimal AO based algorithms for the considered MA schemes to strike a balance between computational complexity and performance.
\vspace{-1.4cm}
\subsection{NOMA}
\vspace{-0.2cm}
As the optimization variables in problem \eqref{P1} are highly-coupled, invoking AO is a common but efficient method for solving such problems. Specifically, we decompose the original problem for a given NOMA decoding order into several subproblems, where the power allocation, IRS reflection coefficients, and deployment location are alternatingly optimized with the other optimization variables being fixed. To reduce the computational complexity caused by exhaustively searching over all possible decoding orders, we further propose an efficient NOMA user ordering scheme based on the user rate weights and IRS-user distances.
\subsubsection{\textbf{Optimizing $\left\{ {{p_k}} \right\}$ for given ${\mathbf{v}}$ and ${\mathbf{s}}$}}
For ease of exposition, we assume that the decoding order is $\mu \left( k \right) = k$ and define ${\beta_k} = \sum\limits_{i = k}^K {{p_i}} ,\forall k \in {\mathcal{K}}$. Therefore, the achievable rate at user $k$ in \eqref{NOMA user k} can be rewritten as
\vspace{-0.3cm}
\begin{align}\label{DC1}
  R_k^N \!=\! {\log _2}\left( {1 \!+\! \frac{{{{\left| {{{\mathbf{q}}_k}{\mathbf{v}}} \right|}^2}{p_k}}}{{{{\left| {{{\mathbf{q}}_k}{\mathbf{v}}} \right|}^2}\sum\nolimits_{i > k}^K {{p_i}}  + {\sigma ^2}}}} \right)\! =\! {\log _2}\left( {{\sigma ^2}\! + \!{{\left| {{{\mathbf{q}}_k}{{\mathbf{v}}}} \right|}^2}{\beta _k}} \right)\! -\! {\log _2}\left( {{\sigma ^2}\! +\! {{\left| {{{\mathbf{q}}_k}{{\mathbf{v}}}} \right|}^2}{\beta _{k + 1}}} \right),
\end{align}
\vspace{-0.8cm}

\noindent where ${\beta _{K + 1}} \triangleq 0$. Therefore, for given ${\mathbf{v}}$ and ${\mathbf{s}}$, the optimization problem can be expressed as
\vspace{-0.3cm}
\begin{subequations}\label{AO-P1}
\begin{align}
\mathop {\max }\limits_{\left\{ {{\beta _k}} \right\}} &\;\;\sum\limits_{k = 1}^K {{w_k}\left( {{{\log }_2}\left( {{\sigma ^2} + {{\left| {{{\mathbf{q}}_k}{{\mathbf{v}}}} \right|}^2}{\beta _k}} \right) - {{\log }_2}\left( {{\sigma ^2} + {{\left| {{{\mathbf{q}}_k}{{\mathbf{v}}}} \right|}^2}{\beta _{k + 1}}} \right)} \right)}     \\
\label{AO-total power1}{\rm{s.t.}}\;\;&{\beta _1} \le {P_{\max }}, {\beta _1} - {\beta _2} \ge {\beta _2} - {\beta _3} \ge  \ldots  \ge {\beta _K} \ge 0.
\end{align}
\end{subequations}
\vspace{-1.2cm}

\noindent Since the objective function is not concave, problem \eqref{AO-P1} is non-convex. However, note that since the objective function is a difference of two concave functions, a concave lower bound can be obtained by applying the first-order Taylor expansion at given local points $ {\beta _{k+1}^{\left( l \right)}} $ in the $l$th iteration of the AO algorithm as follows
\vspace{-0.3cm}
\begin{align}\label{lower bound1}
  R_k^N \!\ge\! \overline R_{k\!\left( {\bm{\beta}}  \right)\!}^N \!=\! {\log _2}\!\left(\! {{\sigma ^2}\! +\! {{\left| {{{\mathbf{q}}_k}{{\mathbf{v}}}} \right|}^2}{\beta _k}} \!\right)\!\!  -\! {\log _2}\!\left(\! {{\sigma ^2} \!+\! {{\left| {{{\mathbf{q}}_k}{{\mathbf{v}}}} \right|}^2}\beta _{k + 1}^{\left( l \right)}} \!\right)\!\! -\! \frac{{{{\left| {{{\mathbf{q}}_k}{{\mathbf{v}}}} \right|}^2}\!\left(\! {{\beta _{k + 1}}\! -\! \beta _{k + 1}^{\left( l \right)}} \!\right)\!{{\log }_2}e}}{{{\sigma ^2}\! + \!{{\left| {{{\mathbf{q}}_k}{{\mathbf{v}}}} \right|}^2}\beta _{k + 1}^{\left( l \right)}}},
\end{align}
\vspace{-0.8cm}

\noindent where ${\bm{\beta}}  = {\left[ {{\beta _1},{\beta _2}, \ldots ,{\beta _K}} \right]^T}$. By replacing the non-concave objective function in problem \eqref{AO-P1} with its concave lower bound, the optimization problem can be written as follows
\vspace{-0.4cm}
\begin{subequations}\label{AO-P2}
\begin{align}
\mathop {\max }\limits_{\left\{ {{\beta _k}} \right\}} &\;\;\sum\limits_{k = 1}^K {{w_k}\overline R _{k\left( {\bm{\beta}} \right)}^N}      \\
\label{AO-total power3}{\rm{s.t.}}\;\;&\eqref{AO-total power1}.
\end{align}
\end{subequations}
\vspace{-1.2cm}

\noindent Now, it can be verified that problem \eqref{AO-P2} is a convex problem which can be efficiently solved via standard convex problem solvers, such as CVX~\cite{cvx}. The objective function values obtained with problem \eqref{AO-P2} in general provide lower bounds for problem \eqref{AO-P1} due to the replacement of the concave lower bound. After solving problem \eqref{AO-P2}, the power allocation can be obtained as ${p_k^*} = {\beta _k^*} - {\beta _{k + 1}^*},\forall k \in {\mathcal{K}}$.
\subsubsection{\textbf{Optimizing ${\mathbf{v}}$ for given $\left\{ {{p_k}} \right\}$ and ${\mathbf{s}}$}} To tackle the non-convex unit modulus constraint, we define ${\mathbf{V}} = {\mathbf{v}}{{\mathbf{v}}^H}$ which satisfies ${{\mathbf{V}}} \succeq 0$, ${\rm {rank}}\left( {{{\mathbf{V}}}} \right) = 1$ and ${\left[ {\mathbf{V}} \right]_{mm}} = 1,m = 1,2, \ldots ,M $, as in the previous section. Then, the achievable rate of user $k$ in \eqref{DC1} can be rewritten as
\vspace{-0.5cm}
\begin{align}\label{DC2}
R_k^N = {\log _2}\left( {{\sigma ^2} + {\rm{Tr}}\left( {{\mathbf{V}}{{\mathbf{Q}}_k}} \right){\beta _k}} \right) - {\log _2}\left( {{\sigma ^2} + {\rm{Tr}}\left( {{\mathbf{V}}{{\mathbf{Q}}_k}} \right){\beta _{k + 1}}} \right),
\end{align}
\vspace{-1.2cm}

\noindent where ${{\mathbf{Q}}_k} = {\mathbf{q}}_k^H{\mathbf{q}}_k^{},\forall k \in {\mathcal{K}}$. As a result, for given $\left\{ {{p_k}} \right\}$ and ${\mathbf{s}}$, the optimization problem can be rewritten as
\vspace{-0.5cm}
\begin{subequations}\label{AO-P3}
\begin{align}
\mathop {\max }\limits_{\left\{ {\mathbf{V}} \right\}} &\;\;\sum\limits_{k = 1}^K {{w_k}\left( {{\log_2}\left( {{\sigma ^2} + {\rm{Tr}}\left( {{\mathbf{V}}{{\mathbf{Q}}_k}} \right){\beta _k}} \right) - {{\log }_2}\left( {{\sigma ^2} + {\rm{Tr}}\left( {{\mathbf{V}}{{\mathbf{Q}}_k}} \right){\beta _{k + 1}}} \right)} \right)}      \\
\label{channel-orderAO}{\rm{s.t.}}\;\;&{{\rm{Tr}}\left( {{\mathbf{V}}{{\mathbf{Q}}_k}} \right)} \ge {{\rm{Tr}}\left( {{\mathbf{V}}{{\mathbf{Q}}_j}} \right)},\forall k > j \in {\mathcal{K}},\\
\label{VmmAO}&{\left[ {\mathbf{V}} \right]_{mm}} = 1,\forall m \in {{\mathcal{M}}},{{\mathbf{V}}}  \succeq  0, {\mathbf{V}} \in {{\mathbb{H}}^{M}},\\
\label{rank 1 VAO}&{\rm {rank}}\left( {{{\mathbf{V}}}} \right) = 1.
\end{align}
\end{subequations}
\vspace{-1.2cm}

\noindent The non-convexity of problem \eqref{AO-P3} is due to the non-concave objective function and the non-convex rank-one constraint. The objective function is the difference of two concave functions, and the concave lower bound based on the first-order Taylor expansion at a given local point ${{\mathbf{V}}^{\left( l \right)}}$ in the $l$th iteration of the AO algorithm is given by
\vspace{-0.5cm}
\begin{align}\label{lower bound2}
\begin{gathered}
  R_k^N \ge \bar R_{k\left( {\mathbf{V}} \right)}^N = {\log _2}\left( {{\sigma ^2} + {\rm{Tr}}\left( {{\mathbf{V}}{{\mathbf{Q}}_k}} \right){\beta _k}} \right) \hfill \\
   - {\log _2}\left( {{\sigma ^2} + {\rm{Tr}}\left( {{{\mathbf{V}}^{\left( l \right)}}{{\mathbf{Q}}_k}} \right){\beta _{k + 1}}} \right) - \frac{{{\beta _{k + 1}}{{\log }_2}e}}{{{\sigma ^2} + {\rm{Tr}}\left( {{{\mathbf{V}}^{\left( l \right)}}{{\mathbf{Q}}_k}} \right){\beta _{k + 1}}}}{\rm{Tr}}\left( {{{\mathbf{Q}}_k}\left( {{\mathbf{V}} - {{\mathbf{V}}^{\left( l \right)}}} \right)} \right). \hfill \\
\end{gathered}
\end{align}
\vspace{-0.8cm}

\noindent Therefore, problem \eqref{AO-P3} can be rewritten as
\vspace{-0.5cm}
\begin{subequations}\label{AO-P4}
\begin{align}
\mathop {\max }\limits_{\left\{ {\mathbf{V}} \right\}} &\;\;\sum\limits_{k = 1}^K {{w_k}\bar R_{k\left( {\mathbf{V}} \right)}^N}       \\
\label{rank 1 VAO2}{\rm{s.t.}}\;\;&{\rm {rank}}\left( {{{\mathbf{V}}}} \right) = 1,\\
\label{con2}&\eqref{channel-orderAO},\eqref{VmmAO}.
\end{align}
\end{subequations}
\vspace{-1.2cm}

\noindent To handle the non-convex rank-one constraint, one common method is to invoke SDR, where the optimization problem is solved by ignoring the rank-one constraint, and then construct a rank-one solution based on the Gaussian randomization method if the solution obtained from the relaxed problem is not rank-one. However, the constructing rank-one solution is generally suboptimal and the convergence of the proposed AO based algorithm may not be guaranteed. To address this issue, we invoke the sequential rank-one
constraint relaxation (SROCR) approach~\cite{SROCR2017}, which is capable of successively finding a rank-one solution. The basic framework of the SROCR approach can be found in \cite{SROCR2017} \textcolor{black}{and its effectiveness has been confirmed in our previous work\footnote{\textcolor{black}{In~\cite{Mu_ax}, we investigated the joint beamforming design in an IRS-assisted multi-antenna NOMA system where the IRS location was assumed to be fixed. We note that the problem formulation and design challenges in~\cite{Mu_ax} are different from those considered in this work.}}~\cite{Mu_ax}.} To facilitate the SROCR approach, the non-convex rank-one constraint \eqref{rank 1 VAO2} is replaced with the following relaxed convex constraint
\vspace{-0.5cm}
\begin{align}\label{sdr}
{\lambda _{\max }}\left( {\mathbf{V}} \right) \ge {\omega  _{\left( i \right)}}{\rm{Tr}}\left( {\mathbf{V}} \right),
\end{align}
\vspace{-1.2cm}

\noindent where ${\lambda _{\max }}\left( {\mathbf{V}} \right)$ denotes the largest eigenvalue of ${\mathbf{V}}$, and ${\omega  _{\left( i \right)}} \in \left[ {0,1} \right]$ denotes a relaxation parameter which controls the largest eigenvalue to trace ratio of ${\mathbf{V}}$. For example, when ${\omega _{\left( i \right)}} = 0$, \eqref{sdr} is equivalent to ignoring the non-convex rank-one constraint as in the SDR approach. When ${\omega _{\left( i \right)}} = 1$, \eqref{sdr} is equivalent to the rank-one constraint. As a result, problem \eqref{AO-P4} can be rewritten as the following relaxed optimization problem
\vspace{-0.3cm}
\begin{subequations}\label{AO-P5}
\begin{align}
\mathop {\max }\limits_{\left\{ {\mathbf{V}} \right\}} &\;\;\sum\limits_{k = 1}^K {{w_k}\bar R_{k\left( {\mathbf{V}} \right)}^N}       \\
\label{rank 1 V P5}{\rm{s.t.}}\;\;&{{\mathbf{u}}_{\max }}{\left( {{{\mathbf{V}}_{\left( i \right)}}} \right)^H}{\mathbf{V}}{{\mathbf{u}}_{\max }}\left( {{{\mathbf{V}}_{\left( i \right)}}} \right) \ge {\omega _{\left( i \right)}}{\rm{Tr}}\left( {\mathbf{V}} \right),\\
\label{con5}&\eqref{channel-orderAO},\eqref{VmmAO},
\end{align}
\end{subequations}
\vspace{-1.2cm}

\noindent where ${{\mathbf{u}}_{\max }}\left( {{{\mathbf{V}}_{\left( i \right)}}} \right)$ is the eigenvector corresponding to the largest eigenvalue of ${{{\mathbf{V}}_{\left( i \right)}}}$ and ${{{\mathbf{V}}_{\left( i \right)}}}$ is the obtained solution feasible with ${\omega _{\left( i \right)}}$ in the $i$th iteration of the SROCR algorithm. Now, problem \eqref{AO-P5} is a convex problem that can be efficiently solved with convex optimization software, such as CVX~\cite{cvx}. \textcolor{black}{After each iteration, parameter ${\omega _{\left( i \right)}}$ is updated as ${\omega _{\left( {i} \right)}}\! =\! \min \! \left(\! {1,\frac{{{{\mathbf{\lambda}} _{\max }}\left( {{{\mathbf{V}}_{\left( {i} \right)}}} \right)}}{{{\rm {Tr}}\left( {{{\mathbf{V}}_{\left( {i} \right)}}} \right)}}\! +\! {\delta _{\left( {i} \right)}}} \right)$, where ${\delta _{\left( i \right)}}$ is a predefined step size. If the current step size ${\delta _{\left( i \right)}}$ makes problem \eqref{AO-P5} infeasible, we reduce the step size as ${\delta _{\left( {i + 1} \right)}}\!\! =\!\! {{{\delta _{\left( i \right)}}} \mathord{\left/
 {\vphantom {{{\delta _{\left( i \right)}}} 2}} \right.
 \kern-\nulldelimiterspace} 2}$. Let ${{\rm {obj}}\!\left( {{{\mathbf{V}}_{\left( i \right)}}} \right)}$ denote the objective function value achieved by solution ${{{\mathbf{V}}_{\left( i \right)}}}$, the algorithm terminates when $\left| {{\rm{obj}}\!\left( {{{\mathbf{V}}_{\left( i \right)}}} \right) \!-\! {\rm{obj}}\!\left( {{{\mathbf{V}}_{\left( {i - 1} \right)}}} \right)} \right| \!\le\! \epsilon_1$ and $\left| {1 - {\omega _{\left( {i - 1} \right)}}} \right| \!\le\! \epsilon_2$ are simultaneously satisfied, where $\epsilon_1$ and $\epsilon_2$ are convergence thresholds. By iteratively solving problem \eqref{AO-P5} and updating parameter ${\omega _{\left( i \right)}}$, the proposed algorithm is guaranteed to converge to a locally optimal rank-one solution~\cite{SROCR2017}. The details for solving problem \eqref{AO-P5} with the SROCR approach are summarized in \textbf{Algorithm 3}.}
\vspace{-0.0cm}
\begin{algorithm}[!t]\label{method3}
\caption{SROCR Approach for Solving Problem \eqref{AO-P5}}
 \hspace*{0.02in}
\hspace*{0.02in}{Initialize convergence thresholds $\epsilon_1$ and $\epsilon_2$, and step size ${\delta _{\left( 0 \right)}}$, and ${\omega _{\left( 0 \right)}} = 0$, $i=0$.}\\
\vspace{-0.4cm}
\begin{algorithmic}[1]
\STATE {\bf repeat}
\STATE For given $\left\{ {{\omega _{\left( i \right)}},{{\mathbf{V}}_{\left( i \right)}}} \right\}$, solve the convex problem \eqref{AO-P5}.
\STATE {\bf if} problem \eqref{AO-P5} is feasible {\bf then}
\STATE ${{{\mathbf{V}}_{\left( {i + 1} \right)}}}={\mathbf{V}}$, ${\delta _{\left( {i + 1} \right)}} = {\delta _{\left( 0 \right)}}$.
\STATE {\bf else}
\STATE ${{{\mathbf{V}}_{\left( {i + 1} \right)}}}={{{\mathbf{V}}_{\left( {i} \right)}}}$, ${\delta _{\left( {i + 1} \right)}} = {{{\delta _{\left( i \right)}}} \mathord{\left/
 {\vphantom {{{\delta _{\left( i \right)}}} 2}} \right.
 \kern-\nulldelimiterspace} 2}$.
\STATE {\bf end}
\STATE $i=i+1$, update ${\omega _{\left( {i} \right)}} = \min \left( {1,\frac{{{{\mathbf{\lambda}} _{\max }}\left( {{{\mathbf{V}}_{\left( {i} \right)}}} \right)}}{{{\rm {Tr}}\left( {{{\mathbf{V}}_{\left( {i} \right)}}} \right)}} + {\delta _{\left( {i} \right)}}} \right)$.
\STATE {\bf until}$\left| {{\rm {obj}}\left( {{{\mathbf{V}}_{\left( i \right)}}} \right) - {\rm {obj}}\left( {{{\mathbf{V}}_{\left( {i - 1} \right)}}} \right)} \right| \le \epsilon_1$ and $\left| {1 - {\omega _{\left( {i - 1} \right)}}} \right| \le \epsilon_2$, ${{{\mathbf{V}}^*} = {{\mathbf{V}}_{\left( i \right)}}}$  .
\end{algorithmic}
\end{algorithm}
\vspace{-0.0cm}
\subsubsection{\textbf{Optimizing ${\mathbf{s}}$ for given ${\mathbf{v}}$ and $\left\{ {{p_k}} \right\}$}} Next, we focus on the optimization of the IRS deployment location for given ${\mathbf{v}}$ and $\left\{ {{p_k}} \right\}$. The achievable rate at user $k$ can be expressed as
\vspace{-0.3cm}
\begin{align}\label{DC3}
R_k^N = {\log _2}\left( {1 + \frac{{{{\left| {\overline {\mathbf{q}} _k{\mathbf{v}}} \right|}^2}{p_k}}}{{\sum\nolimits_{k > i}^K {{{\left| {\overline {\mathbf{q}} _k{\mathbf{v}}} \right|}^2}{p_i}}  + \frac{{{\sigma ^2}}}{{{L_{IRS,k}}}}}}} \right),
\end{align}
\vspace{-0.8cm}

\noindent where $\overline {\mathbf{q}}_k =  {{\mathbf{r}}_k^H}{\rm{diag}}\left( {\mathbf{g}} \right)$. Note that since the path loss ${{L_{IRS,k}}}$ and the AoAs/AoDs in ${\overline {\mathbf{q}} _k}$ both depend on the IRS deployment location, the optimization of the IRS deployment location is a non-trivial task. To handle this difficulty, we propose a \emph{local region optimization} method for the design of the IRS deployment location. For a given feasible IRS deployment location ${{{\mathbf{s}}^{\left( l \right)}}}$, the optimized IRS location should satisfy the following condition
\vspace{-0.5cm}
\begin{align}\label{delta}
\left\| {{\mathbf{s}} - {{\mathbf{s}}^{\left( l \right)}}} \right\| \le \Delta,
\end{align}
\vspace{-1.2cm}

\noindent where $\Delta $ is chosen to be relatively small such that the AoAs/AoDs of the AP-IRS and IRS-user links remain approximately unchanged during the current iteration. Therefore, ${\left| {{{\overline {\mathbf{q}} }_k}{\mathbf{v}}} \right|^2}$ is approximately constant. The choice of $\Delta $ will be discussed in Section VI.\\
\indent For given $\left\{ {{p_k}} \right\}$, ${\mathbf{v}}$, and ${{{\mathbf{s}}^{\left( l \right)}}}$, the optimization problem within a specific local region can be expressed as
\vspace{-0.5cm}
\begin{subequations}\label{AO-P6}
\begin{align}
\mathop {\max }\limits_{\mathbf{s}} &\;\;\sum\limits_{k = 1}^K {{w_k}{{\log }_2}\left( {1 + \frac{{{{\overline c }_k}{p_k}}}{{\sum\nolimits_{i > k}^K {{{\overline c }_k}{p_i}}  + \frac{{{\sigma ^2}}}{{{L_{IRS,k}}}}}}} \right)} \\
\label{deploymentP632}{\rm{s.t.}}\;\;&\frac{{\rho _0^{}}}{{d_{AI}^{{\alpha _{AI}}}}}\frac{{\rho _0^{}}}{{d_{IU,k}^{{\alpha _{IU}}}}}{{\bar c}_k} \ge \frac{{\rho _0^{}}}{{d_{AI}^{{\alpha _{AI}}}}}\frac{{\rho _0^{}}}{{d_{IU,j}^{{\alpha _{IU}}}}}{{\bar c}_j},\forall k > j,\\
\label{deploymentP6}&{\mathbf{s}} \in \Omega ,\left\| {{\mathbf{s}} - {{\mathbf{s}}^{\left( l \right)}}} \right\| \le \Delta,
\end{align}
\end{subequations}
\vspace{-1.2cm}

\noindent where ${\overline c _k} \triangleq {\left| {{{\overline {\mathbf{q}} }_k}{\mathbf{v}}} \right|^2}, \forall k \in {\mathcal{K}}$. Constraint \eqref{deploymentP632} is the NOMA decoding order constraint for the IRS deployment location design problem, which can be equivalently expressed as
\vspace{-0.5cm}
\begin{align}\label{47}
{{\overline c}_k}^{\frac{2}{{{\alpha _{IU}}}}}{\left\| {{\mathbf{s}} - {{\mathbf{u}}_j}} \right\|^2} \ge {{\overline c}_j}^{\frac{2}{{{\alpha _{IU}}}}}{\left\| {{\mathbf{s}} - {{\mathbf{u}}_k}} \right\|^2},\forall k > j,
\end{align}
\vspace{-1.2cm}

\noindent where ${{\mathbf{u}}_j}$ and ${{\mathbf{u}}_j}$ denote the location of users $j$ and $k$, respectively.\\
\indent Next, we introduce auxiliary variables $\varphi $, $\left\{ {{\upsilon _k}} \right\}$, and $\left\{ {{\tau _k}} \right\}$ such that
\vspace{-0.5cm}
\begin{align}
\label{aa1}\varphi  &= {\left\| {{\mathbf{s}} - {\mathbf{b}}} \right\|^{{\alpha _{AI}}}},\\
\label{aa2}{\upsilon _k} &= {\left\| {{\mathbf{s}} - {{\mathbf{u}}_k}} \right\|^{{\alpha _{IU}}}},\forall k,\\
\label{aa3}{\tau _k} &= \frac{1}{{{L_{IRS,k}}}} = \frac{{\varphi {\upsilon _k}}}{{\rho _0^2}},\forall k.
\end{align}
\vspace{-1cm}

\noindent Then, problem \eqref{AO-P6} can be equivalently written as
\vspace{-0.3cm}
\begin{subequations}\label{AO-P7}
\begin{align}
\mathop {\max }\limits_{{\mathbf{s}},\varphi ,\left\{ {{\upsilon _k}} \right\},\left\{ {{\tau _k}} \right\}} {\text{ }}&\;\;\sum\limits_{k = 1}^K {{w_k}{{\log }_2}\left( {1 + \frac{{{{\bar c}_k}{p_k}}}{{\sum\nolimits_{i > k}^K {{{\bar c}_k}{p_i}}  + {\tau _k}{\sigma ^2}}}} \right)}  \\
\label{deploymentP63}{\rm{s.t.}}\;\;&{\tau _k} \ge \frac{{\varphi {\upsilon _k}}}{{\rho _0^2}},\forall k \in {\mathcal{K}},\\
\label{deploymentP64}&\varphi  \ge {\left\| {{\mathbf{s}} - {\mathbf{b}}} \right\|^{{\alpha _{AI}}}},\\
\label{deploymentP65}&{\upsilon _k} \ge {\left\| {{\mathbf{s}} - {{\mathbf{u}}_k}} \right\|^{{\alpha _{IU}}}},\forall k \in {{\mathcal{K}}},\\
\label{deploymentP66}&\eqref{deploymentP6},\eqref{47}.
\end{align}
\end{subequations}
\vspace{-1.2cm}

\noindent \textcolor{black}{Note that in problem \eqref{AO-P7}, constraints \eqref{deploymentP63}, \eqref{deploymentP64}, and \eqref{deploymentP65} are obtained from \eqref{aa1}, \eqref{aa2}, and \eqref{aa3} by replacing the equality sign with the
inequality sign. This does not affect the equivalence of problems \eqref{AO-P7} and \eqref{AO-P6}. To demonstrate this, assume that one of the corresponding constraints in \eqref{deploymentP63} is satisfied with strict inequality. Then, we can always decrease the corresponding ${\tau _k}$ to satisfy the constraint with equality, which in turn increases the objective value. Therefore, for the optimal solution of problem \eqref{AO-P7}, constraint \eqref{deploymentP63} must be satisfied with equality. Similarly, constraints \eqref{deploymentP64} and \eqref{deploymentP65} also must be satisfied with equality at the optimal solution of problem \eqref{AO-P7}.} Now, the non-convexity of problem \eqref{AO-P7} is caused by the non-concave objective function and the non-convex constraints \eqref{47} and \eqref{deploymentP63}. Since the objective function is a convex function with respect to ${{\tau _k}}$, a lower bound based on the first-order Taylor expansion at a given point $\tau _k^{\left( l \right)}$ in the $l$th iteration of the AO algorithm is given by
\vspace{-0.4cm}
\begin{align}\label{lower bound3}
  R_k^N\! \ge\! \overline R_{k\left( \tau  \right)}^N \!=\! {\log _2}\left( {\!1 \!+\! \frac{{{{\overline c}_k}{p_k}}}{{\sum\nolimits_{i > k}^K {{{\overline c}_k}{p_i}}\!  + \!\tau _k^{\left( l \right)}{\sigma ^2}}}} \right)  - \frac{{{{\overline c}_k}{p_k}{\sigma ^2}{{\log }_2}e\left( {\tau _k\! -\! \tau _k^{\left( l \right)}} \right)}}{{\left( {\sum\nolimits_{i > k}^K {{{\overline c}_k}{p_i}} \! + \!\tau _k^{\left( l \right)}{\sigma ^2}\! +\! {{\overline c}_k}{p_k}} \right)\!\!\left( {\sum\nolimits_{i > k}^K {{{\overline c}_k}{p_i}}\!  +\! \tau _k^{\left( l \right)}{\sigma ^2}} \right)}}.
\end{align}
\vspace{-0.8cm}

\noindent For the non-convex constraint \eqref{47}, as ${\left\| {{\mathbf{s}} - {{\mathbf{u}}_j}} \right\|^2}$ is a convex function
with respect to ${\mathbf{s}}$, a lower bound based on the first-order Taylor expansion at a given point ${{{\mathbf{s}}^{\left( l \right)}}}$ is given by
\vspace{-0.4cm}
\begin{align}\label{SS}
{\left\| {{\mathbf{s}} - {{\mathbf{u}}_j}} \right\|^2} \ge {\left\| {{{\mathbf{s}}^{\left( l \right)}} - {{\mathbf{u}}_j}} \right\|^2} + 2{\left( {{{\mathbf{s}}^{\left( l \right)}} - {{\mathbf{u}}_j}} \right)^T}\left( {{\mathbf{s}} - {{\mathbf{s}}^{\left( l \right)}}} \right).
\end{align}
\vspace{-1.2cm}

\noindent To handle non-convex constraint \eqref{deploymentP63}, the left-hand-side (LHS) can be expressed as
\vspace{-0.3cm}
\begin{align}\label{DC4}
\varphi {\upsilon _k} = \frac{{{{\left( {\varphi  + {\upsilon _k}} \right)}^2}}}{2} - \frac{1}{2}{\varphi ^2} - \frac{1}{2}\upsilon _k^2.
\end{align}
\vspace{-1.1cm}

\noindent By applying the first-order Taylor expansion, the convex lower bound at given local points ${\varphi ^{\left( l \right)}}$ and $\left\{ {\upsilon _k^{\left( l \right)}} \right\}$ is given by
\vspace{-0.5cm}
\begin{align}\label{lower bound4}
  \varphi {\upsilon _k} \ge {\varpi _k} = \frac{{{{\left( {\varphi  + {\upsilon _k}} \right)}^2}}}{2}    - \left( {\frac{1}{2}{\varphi ^{\left( l \right)}}^2 + {\varphi ^{\left( l \right)}}\left( {\varphi  - {\varphi ^{\left( l \right)}}} \right)} \right) - \left( {\frac{1}{2}\upsilon _k^{\left( l \right)2} + \upsilon _k^{\left( l \right)}\left( {\upsilon _k^{} - \upsilon _k^{\left( l \right)}} \right)} \right).
\end{align}
\vspace{-1.1cm}

\noindent Exploiting \eqref{lower bound3}, \eqref{SS}, and \eqref{lower bound4}, problem \eqref{AO-P7} can be rewritten as
\vspace{-0.4cm}
\begin{subequations}\label{AO-P8}
\begin{align}
&\mathop {\max }\limits_{{\mathbf{s}},\varphi ,\left\{ {{\upsilon _k}} \right\},\left\{ {{\tau _k}} \right\}} \;\;\sum\limits_{k = 1}^K {{w_k}\bar R_{k\left( \tau  \right)}^N}   \\
\label{deploymentP69}{\rm{s.t.}}\;\;&{\left\| {{{\mathbf{s}}^{\left( l \right)}} - {{\mathbf{u}}_j}} \right\|^2} + 2{\left( {{{\mathbf{s}}^{\left( l \right)}} - {{\mathbf{u}}_j}} \right)^T}\left( {{\mathbf{s}} - {{\mathbf{s}}^{\left( l \right)}}} \right) \ge {\left( {\frac{{{{\overline c}_j}}}{{{{\overline c}_k}}}} \right)^{\frac{2}{{{\alpha _{IU}}}}}}{\left\| {{\mathbf{s}} - {{\mathbf{u}}_k}} \right\|^2},\\
\label{deploymentP67}&{\tau _k} \ge \frac{{{\varpi _k}}}{{\rho _0^2}},\forall k \in {{\mathcal{K}}},\\
\label{deploymentP68}&\eqref{deploymentP6},\eqref{deploymentP64},\eqref{deploymentP65}.
\end{align}
\end{subequations}
\vspace{-1.2cm}

\noindent Problem \eqref{AO-P8} is convex and can be efficiently solved with convex optimization software, such as CVX~\cite{cvx}. Similarly, the obtained objective value serves as a lower bound for that of the original problem \eqref{AO-P7} because of the bounds in \eqref{lower bound3}, \eqref{SS}, and \eqref{lower bound4}.
\subsubsection{\textbf{Proposed NOMA User Ordering Scheme}} Note that the above subproblems are solved for a given decoding order. A straightforward approach is to exhaustively search over all possible decoding orders and to choose the best candidate solution. However, this approach requires a prohibitive complexity, e.g., ${{\mathcal{O}}}\left( {K!} \right)$, which is unacceptable even for moderate values of $K$. To address this issue, we propose a user ordering scheme based on the user rate weights and the IRS-user link distances. \textcolor{black}{To maximize the weighted sum rate of all users, one intuitive solution is to allocate higher data rates to the users with higher rate weights. Motivated by this, the decoding orders of the users are firstly differentiated based on the user rate weights as follows:}
\vspace{-0.5cm}
\begin{align}
\textcolor{black}{\mu \left( k \right) < \mu \left( j \right)\;\;{\rm{if}}\;{w_k} < {w_j},\forall k,j \in {\mathcal{K}}.}
\end{align}
\vspace{-1.2cm}

\noindent \textcolor{black}{In this way, users with higher rate weights suffer less interference from the other users. Moreover, the decoding orders of users who have identical rate weights are further determined based on the distances between the users and the initial IRS deployment location ${{{\mathbf{s}}^{\left( 0 \right)}}}$ as follows:}
\vspace{-0.5cm}
\begin{align}
\textcolor{black}{\mu \left( k \right) < \mu \left( j \right)\;\;{\rm{if}}\;{w_k} = {w_j}\;{\rm{and}}\;\left\| {{\mathbf{s}^{\left( 0 \right)}} - {{\mathbf{u}}_j}} \right\| \le \left\| {{\mathbf{s}^{\left( 0 \right)}} - {{\mathbf{u}}_k}} \right\|,\forall k,j \in {\mathcal{K}},}
\end{align}
\vspace{-1.2cm}

\noindent \textcolor{black}{which means that users close to the IRS have higher decoding orders. The performance of the proposed user ordering scheme will be compared with the exhaustive search in the numerical results section.}
\subsubsection{\textbf{Proposed Algorithm and Convergence}} Based on the above three subproblems and the proposed user ordering scheme, we design an AO based algorithm. In particular, the power allocation, $\left\{ {{p_k}} \right\}$, the IRS reflection coefficients, ${\mathbf{v}}$, and the IRS deployment location, ${\mathbf{s}}$, are alternately designed by solving subproblems \eqref{AO-P2}, \eqref{AO-P4}, and \eqref{AO-P8}, respectively. The solutions obtained in each iteration are used as the input local points for the next iteration. The details of the proposed algorithm are summarized in \textbf{Algorithm 4}, where $N_s>1$ candidate local region points ${{\mathbf{s}}^{\left( 0 \right)}}$ are initialized for optimization, from which the solution with the highest WSR is selected as the final solution. In particular, for each initialized IRS deployment location, the user decoding order is generated based on the proposed user ordering scheme, and the transmit power is equally distributed among all users for randomly generated feasible IRS reflection coefficients.\\
\renewcommand*\footnoterule{}
\begin{savenotes}
\begin{algorithm}[!t]\label{method4}
\caption{Proposed AO based Algorithm for Solving Problem \eqref{P1} for NOMA}
 \hspace*{0.02in}
 \hspace*{0.02in}{Initialize $N_s$ candidate IRS deployment locations $\left\{ {{\mathbf{s}}_n^{^{\left( 0 \right)}}} \right\}_{n = 1}^{{N_s}}$.}\\
\vspace{-0.4cm}
\begin{algorithmic}[1]
\STATE {\bf for} $n=1$ to $N_s$ {\bf do}
\STATE $l=0$, initialize the decoding order of all users according to the user rate weights and IRS-user distances, equal power allocation $\left\{ {p_{k,n}^{\left( l \right)}  = \frac{{{P_{\max }}}}{K}} \right\}$, and random feasible IRS reflection coefficients ${{\mathbf{v}}_n^{\left( l \right)}}$.
\STATE {\bf repeat}
\STATE Solve problem \eqref{AO-P2} once\footnote{\textcolor{black}{We note that problem \eqref{AO-P2} can also be solved iteratively until convergence. To strike a balance between performance and complexity, we solve problem \eqref{AO-P2} only once.}} for given ${\mathbf{v}}_n^{\left( l \right)}$ and ${{\mathbf{s}}_n^{\left( l \right)}}$, and denote the optimal solution by $\left\{ {{{p}}_{k,n}^{\left( l+1 \right)}} \right\}$.
\STATE Solve problem \eqref{AO-P5} iteratively with the proposed \textbf{Algorithm 3} for given $\left\{ {{{p}}_{k,n}^{\left( l+1 \right)}} \right\}$ and ${{\mathbf{s}}_n^{\left( l \right)}}$, and denote the optimal solution by ${\mathbf{v}}_n^{\left( l+1 \right)}$.
\STATE Calculate ${\varphi _{n}^{\left( l \right)}},\left\{ {\upsilon _{k,n}^{\left( l \right)}} \right\},\left\{ {\tau _{k,n}^{\left( l \right)}} \right\}$ based on \eqref{aa1}, \eqref{aa2}, and \eqref{aa3}.
\STATE Solve problem \eqref{AO-P8} for given $\left\{ {{{p}}_{k,n}^{\left( l+1 \right)}} \right\}$ and ${{\mathbf{v}}_n^{\left( l \right)}}$ in the local region of ${{\mathbf{s}}_n^{\left( l \right)}}$, and denote the optimal solution by ${\mathbf{s}}_n^{\left( l+1 \right)}$.
\STATE $l=l+1$.
\STATE {\bf until} the fractional increase of the objective value is below a threshold $\xi   > 0$
\STATE Record the $n$th obtained solutions ${{\rm T}_n} = \left( {\left\{ {p_{n,k}^{\left( l \right)}} \right\},{\mathbf{v}}_n^{\left( l \right)},{\mathbf{s}}_n^{\left( l \right)}} \right)$ and the corresponding objective function value ${{\rm{WSR}}_n} = \sum\nolimits_{k = 1}^K {{w_k}R_{n,k}^{N*}} $.
\STATE {\bf end}
\STATE Select the desired solutions ${{\rm T}_{{n^*}}}$, where ${n^*} = \mathop {\arg \max }\limits_{n = 1, \ldots ,{N_s}} \;{{\rm{WSR}}_n}$.
\end{algorithmic}
\end{algorithm}
\end{savenotes}
\indent To prove the convergence of Algorithm 4, let $\eta \left( {\left\{ {p_{k,n}^l} \right\},{{\mathbf{v}}_n^l},{{\mathbf{s}}_n^l}} \right)$ denote the objective function value of problem \eqref{P1} in the $l$th iteration with the $n$th candidate IRS deployment location. First, for power allocation optimization for given ${{\mathbf{v}}_n^l}$ and ${{\mathbf{s}}_n^l}$ in step 4 of Algorithm 4, we have
\vspace{-0.4cm}
\begin{align}\label{c1}
  \textcolor{black}{\eta\! \left( {\left\{ {p_{k,n}^l} \right\}\!,\!{{\mathbf{v}}_n^l},{{\mathbf{s}}_n^l}} \right)\!\mathop  = \limits^{\left( a \right)} \!\eta _{\left\{ {p_{k,n}^l} \right\}}^{lb}\!\left( {\left\{ {p_{k,n}^l} \right\}\!,\!{{\mathbf{v}}_n^l},{{\mathbf{s}}_n^l}} \right)\! \mathop  \le \limits^{\left( b \right)}\! \eta _{\left\{ {p_{k,n}^l} \right\}}^{lb}\!\left( {\left\{ {p_{k,n}^{l + 1}} \right\}\!,\!{{\mathbf{v}}_n^l},{{\mathbf{s}}_n^l}} \right) \!  \mathop \le \limits^{\left( c \right)} \!\eta\! \left( {\left\{ {p_{k,n}^{l + 1}} \right\}\!,\!{{\mathbf{v}}_n^l},{{\mathbf{s}}_n^l}} \right)\!,\!}
\end{align}
\vspace{-1.2cm}

\noindent \textcolor{black}{where $\eta _{\left\{ {p_{k,n}^l} \right\}}^{lb}$ denotes the objective function value of problem \eqref{AO-P2} for the local points ${\left\{ {p_{k,n}^l} \right\}}$.} Here, ($a$) follows from the fact that the first-order Taylor expansion in \eqref{lower bound1} is tight at the given local point; ($b$) holds since solution ${\left\{ {p_{k,n}^{l + 1}} \right\}}$ for problem \eqref{AO-P2} is optimal for given ${{\mathbf{v}}^l}$ and ${{\mathbf{s}}^l}$; ($c$) is obtained since problem \eqref{AO-P2} always provides a lower bound solution for the original problem \eqref{P1}. \textcolor{black}{Therefore, according to \eqref{c1}, for fixed ${{\mathbf{v}}_n^l}$ and ${{\mathbf{s}}_n^l}$, the objective value of (9) is non-decreasing after solving power allocation subproblem (36).} Similarly, for the IRS reflection coefficient and deployment location optimization in step 5 and step 7 of Algorithm 4, it can be shown that
\vspace{-0.5cm}
\begin{align}\label{c2}
  \textcolor{black}{\eta \left( {\left\{ {p_{k,n}^{l + 1}} \right\},{{\mathbf{v}}_n^l},{{\mathbf{s}}_n^l}} \right) \!= \!\eta _{{\mathbf{v}}_n^l}^{lb}\left( {\left\{ {p_{k,n}^{l + 1}} \right\},{{\mathbf{v}}_n^l},{{\mathbf{s}}_n^l}} \right) \! \le \! \eta _{{\mathbf{v}}_n^l}^{lb}\left( {\left\{ {p_{k,n}^{l + 1}} \right\},{{\mathbf{v}}_n^{l + 1}},{{\mathbf{s}}_n^l}} \right) \! \le \! \eta \left( {\left\{ {p_{k,n}^{l + 1}} \right\},{{\mathbf{v}}_n^{l + 1}},{{\mathbf{s}}_n^l}} \right)}
\end{align}
\vspace{-1.2cm}

\noindent and
\vspace{-0.5cm}
\begin{align}\label{c3}
  \textcolor{black}{\eta \left( {\left\{ {p_{k,n}^{l + 1}} \right\}\!,\!{{\mathbf{v}}_n^{l + 1}}\!,\!{{\mathbf{s}}_n^l}} \right)\! = \!\eta _{\mathbf{s}_n^l}^{lb}\left( {\left\{ {p_{k,n}^{l + 1}} \right\}\!,\!{{\mathbf{v}}_n^{l + 1}}\!,\!{{\mathbf{s}}_n^l}} \right)\! \le\! \eta _{\mathbf{s}_n^l}^{lb}\left( {\left\{ {p_{k,n}^{l + 1}} \right\}\!,\!{{\mathbf{v}}_n^{l + 1}}\!,\!{{\mathbf{s}}_n^{l + 1}}} \right)\!  \le\! \eta \left( {\left\{ {p_{k,n}^{l + 1}} \right\}\!,\!{{\mathbf{v}}_n^{l + 1}}\!,\!{{\mathbf{s}}_n^{l + 1}}} \right),}
\end{align}
\vspace{-1.2cm}

\noindent \textcolor{black}{where $\eta _{\mathbf{v}_n^l}^{lb}$ and $\eta _{\mathbf{s}_n^l}^{lb}$ denote the objective function values of subproblems \eqref{AO-P4} and \eqref{AO-P8} for the local points ${{\mathbf{v}_n^l}}$ and ${{\mathbf{s}_n^l}}$, respectively.}\\
\indent Therefore, based on \eqref{c1}-\eqref{c3}, we have
\vspace{-0.5cm}
\begin{align}\label{c4}
\eta \left( {\left\{ {p_{k,n}^l} \right\},{{\mathbf{v}}_n^l},{{\mathbf{s}}_n^l}} \right) \le \eta \left( {\left\{ {p_{k,n}^{l + 1}} \right\},{{\mathbf{v}}_n^{l + 1}},{{\mathbf{s}}_n^{l + 1}}} \right).
\end{align}
\vspace{-1.6cm}
\begin{remark}\label{convegence}
\emph{Equation \eqref{c4} shows that, \textcolor{black}{by alternatingly optimizing the three types of optimization variables,} the obtained value of problem \eqref{P1} is non-decreasing in each iteration of \textbf{Algorithm 4}. Since the WSR is upper bounded by a finite value, the proposed algorithm is guaranteed to converge.}
\vspace{-0.5cm}
\begin{remark}\label{SDR4}
\emph{\textcolor{black}{Similarly, for the obtained IRS deployment location ${\mathbf{s}}_{n^{*}}$, the power allocation and the IRS reflection coefficients can be alternatingly optimized again based on I-CSI with \textbf{Algorithm 4} to find a suboptimal solution.}}
\end{remark}
\end{remark}
\vspace{-1cm}
\subsection{OMA}
\vspace{-0.2cm}
In this subsection, we propose AO based algorithms for solving the formulated optimization problems for FDMA and TDMA.
\subsubsection{\textbf{FDMA}} For given IRS reflection coefficients and deployment locations, the power allocation optimization problem for FDMA can be expressed as
\vspace{-0.4cm}
\begin{subequations}\label{P2-1}
\begin{align}
\mathop {\max }\limits_{\left\{ {{p_k}} \right\}} &\;\;\sum\limits_{k = 1}^K {{w_k}\frac{1}{K}{{\log }_2}\left( {1 + \frac{{{{\left| {{{\mathbf{q}}_k}{{\mathbf{v}}}} \right|}^2}{p_k}}}{{\frac{1}{K}{\sigma ^2}}}} \right)}  \\
\label{P2-1 con}{\rm{s.t.}}\;\;&{p_k}\ge 0,\forall k \in {\mathcal{K}}, \sum\limits_{k = 1}^K {{p_k}}  \le P_{\max}.
\end{align}
\end{subequations}
\vspace{-1cm}

\noindent Since the objective function is concave with respect to ${\left\{ {{p_k}} \right\}}$, problem \eqref{P2-1} is convex and can be solved via standard convex problem solvers such as CVX~\cite{cvx}.\\
\indent Next, for given power allocation and IRS deployment location, the IRS reflection coefficient optimization problem can be written as
\vspace{-0.4cm}
\begin{subequations}\label{P2-2}
\begin{align}
\mathop {\max }\limits_{\left\{ {\mathbf{V}} \right\}} &\;\;\sum\limits_{k = 1}^K {{w_k}\frac{1}{K}{{\log }_2}\left( {1 + \frac{{{\rm{Tr}}\left( {{\mathbf{V}}{{\mathbf{Q}}_k}} \right){p_k}}}{{\frac{1}{K}{\sigma ^2}}}} \right)}       \\
\label{P2-2 VmmAO}{\rm{s.t.}}\;\;&{\left[ {\mathbf{V}} \right]_{mm}} = 1,\forall m \in {{\mathcal{M}}},{{\mathbf{V}}}  \succeq  0, {\mathbf{V}} \in {{\mathbb{H}}^{M}},\\
\label{P2-2 rank 1 VAO}&{\rm {rank}}\left( {{{\mathbf{V}}}} \right) = 1,
\end{align}
\end{subequations}
\vspace{-1.2cm}

\noindent where ${\mathbf{V}} = {\mathbf{v}}{{\mathbf{v}}^H}$, as in the previous subsection. The non-convexity of problem \eqref{P2-2} is caused by non-convex rank-one constraint \eqref{P2-2 rank 1 VAO}. Similar to problem \eqref{AO-P5}, problem \eqref{P2-2} also satisfies the general framework of the SROCR approach. Therefore, problem \eqref{P2-2} can be solved with the SROCR approach via \textbf{Algorithm 3}. The details are omitted here for brevity.\\
\indent For the IRS deployment location optimization, we invoke again the proposed local region optimization method. As a result, for given $\left\{ {{p_k}} \right\}$, ${\mathbf{v}}$, and local region point ${{{\mathbf{s}}^{\left( l \right)}}}$, the optimization problem is given by
\vspace{-0.4cm}
\begin{subequations}\label{P2-3}
\begin{align}
\mathop {\max }\limits_{\mathbf{s}} &\;\;\sum\limits_{k = 1}^K {{w_k}\frac{1}{K}{{\log }_2}\left( {1 + \frac{{{{\overline c}_k}{p_k}}}{{\frac{1}{K}\frac{{{\sigma ^2}}}{{{L_{IRS,k}}}}}}} \right)}  \\
\label{P2-3 deploymentP6}{\rm{s.t.}}\;\;&{\mathbf{s}} \in \Omega ,\left\| {{\mathbf{s}} - {{\mathbf{s}}^{\left( l \right)}}} \right\| \le \Delta.
\end{align}
\end{subequations}
\vspace{-1.2cm}

\noindent It is observed that problem \eqref{P2-3} has a similar structure as problem \eqref{AO-P6} for NOMA. Therefore, problem \eqref{P2-3} can be solved in the same manner as problem \eqref{AO-P8}.
\subsubsection{\textbf{TDMA}} For TDMA, we only need to optimize the IRS reflection coefficients and deployment locations. The design the of reflection coefficients for given IRS location has already been addressed in problem \eqref{P3 sub} in Section III. Next, we focus on the IRS deployment location design. By invoking the local region optimization method, the corresponding optimization problem is given by
\vspace{-0.4cm}
\begin{subequations}\label{P3-1}
\begin{align}
\mathop {\max }\limits_{\mathbf{s}} &\;\;\sum\limits_{k = 1}^K {{w_k}\frac{1}{K}{{\log }_2}\left( {1 + \frac{{{{\bar c}_k}{P_{\max}}}}{{\frac{{{\sigma ^2}}}{{{L_{IRS,k}}}}}}} \right)}  \\
\label{P3-1 deploymentP6}{\rm{s.t.}}\;\;&{\mathbf{s}} \in \Omega ,\left\| {{\mathbf{s}} - {{\mathbf{s}}^{\left( l \right)}}} \right\| \le \Delta.
\end{align}
\end{subequations}
\vspace{-1.2cm}

\noindent Problem \eqref{P3-1} can be efficiently solved by the SCA method as described in the previous subsection. The details are omitted here for brevity.
\vspace{-0.6cm}
\section{Discussion of Complexity and Performance}
\vspace{-0.2cm}
In this section, we discuss the computational complexity and achieved performance of the proposed algorithms. The overall comparison is summarized in Table \ref{comparison}. \textcolor{black}{Specifically, ``MO-EX-NOMA'' denotes the proposed MO based algorithm for NOMA in Section III-A, where the user decoding order and IRS deployment location are obtained by an exhaustive search. ``MO-EX-FDMA'' denotes the corresponding MO based algorithm for FDMA in Section III-B. In ``EX-TDMA'', the optimal IRS reflection coefficients are designed based on the proposed closed-form solution and the IRS deployment location is obtained by an exhaustive search for TDMA, see Section III-C. ``AO-NOMA'', ``AO-FDMA'', and ``AO-TDMA'' denote the proposed AO based algorithms for the different MA schemes in Section IV.}\\
\indent We first analyze the complexity of the ``MO-EX-NOMA'' algorithm for solving problem \eqref{P1}. The complexity of exhaustively searching all possible decoding orders and IRS deployment locations over the 3D space with accuracy $\xi $ are ${\mathcal{O}}\left( {K!} \right)$ and ${\mathcal{O}}\left( {\frac{1}{{{\xi ^3}}}} \right)$, respectively. For the power allocation and IRS reflection coefficient optimization, the main complexity originates from finding the projection with Algorithm 2. The complexity of solving the semidefinite programming (SDP) problem \eqref{check} in Algorithm 2 is ${\mathcal{O}}\left( {\max {{\left( {M,2K - 1} \right)}^4}\sqrt M } \right)$~\cite{Luo}. The total complexity of Algorithm 2 with accuracy $\varepsilon $ is ${\mathcal{O}}\left( {{{\log }_2}\frac{1}{\varepsilon }\left( {\max {{\left( {M,2K - 1} \right)}^4}\sqrt M } \right)} \right)$. Let $I_{ite}^M$ denote the number of iterations needed for the convergence of Algorithm 1, then the overall complexity of the ``MO-EX-NOMA'' algorithm is as shown in Table \ref{comparison}. Similarly, we can also obtain the complexities of the ``MO-EX-FDMA'' and ``EX-TDMA'' algorithms as shown in Table \ref{comparison}.\\
\indent Next, we analyze the complexity of ``AO-NOMA''. Since the user decoding order is generated with the proposed user ordering scheme, its complexity is ${\mathcal{O}}\left( 1 \right)$. Assuming application of the interior-point method, the complexities of the power allocation and IRS deployment location design are ${\mathcal{O}}\left( {{K^{3.5}}} \right)$ and ${\mathcal{O}}\left( {{{\left( {2K + 4} \right)}^{3.5}}} \right)$, respectively~\cite{convex}. For the IRS reflection coefficient design, the complexity of Algorithm 3 is ${\mathcal{O}}\left( {I_{ite}^S\left( {\max {{\left( {M,2K - 1} \right)}^4}\sqrt M } \right)} \right)$, where $I_{ite}^S$ denotes the number of iterations needed for the convergence of Algorithm 3. The total complexity of one iteration in Algorithm 4 is ${\mathcal{O}}\left( {{{\left( {3K + 4} \right)}^{3.5}} + I_{ite}^S\left( {\max {{\left( {M,2K - 1} \right)}^4}\sqrt M } \right)} \right)$. Let $I_{ite}^A$ denote the number of iterations needed for the convergence of Algorithm 4, then the overall complexity of the ``AO-NOMA'' algorithm is as shown in Table \ref{comparison}. The complexity of ``AO-FDMA'' and ``AO-TDMA'' can also be analyzed in a similar manner.\\
\begin{table*}[!t]\tiny
\caption{Computational Complexity and Performance of Proposed Algorithms.}
\vspace{-0.6cm}
\begin{center}
\centering
\resizebox{\textwidth}{!}{
\begin{tabular}{|l|l|l|l|l|l|l|}
\hline
\centering
 \textbf{\makecell[c]{Proposed \\Algorithm}} &\textbf{\makecell[c]{Decoding \\ order}}  & \textbf{\makecell[c]{Power \\allocation}} & \textbf{\makecell[c]{Reflection \\coefficients}} &\textbf{\makecell[c]{Deployment \\location}}& \textbf{\makecell[c]{Complexity}} & \textbf{\makecell[c]{Performance}}\\
\hline
\centering
MO-EX-NOMA & \makecell[c]{Exhaustive \\ search} & \multicolumn{2}{c|}{\makecell[c]{Monotonic\\ optimization}} & \makecell[c]{Exhaustive\\ search} & ${\mathcal{O}}\left( {\frac{{K!}}{{{{\xi }^3}}}\left( {I_{ite}^{M}{{\log }_2}\frac{1}{{\varepsilon}}\left( {\max \left( {M,2K - 1} \right)^4\sqrt M} \right)} \right)} \right)$& \makecell[c]{Upper\\ bound}\\
\hline
\centering
\makecell[c]{MO-EX-FDMA} & \diagbox{\qquad \qquad}{\qquad \qquad}  & \multicolumn{2}{c|}{\makecell[c]{Monotonic\\ optimization}} & \makecell[c]{Exhaustive \\ search} & ${\mathcal{O}}\left( {\frac{1}{{{{\xi}^3}}}\left( {I_{ite}^{M}{{\log }_2}\frac{1}{{\varepsilon }}\left( {\max \left( {M,K} \right)^4\sqrt M } \right)} \right)} \right)$& \makecell[c]{Upper\\ bound}\\
\hline
\centering
\makecell[c]{EX-TDMA} & \diagbox{\qquad \qquad}{\qquad \qquad} & \diagbox{\qquad \qquad}{\qquad \qquad} & \makecell[c]{Closed-form\\ solution} & \makecell[c]{Exhaustive \\ search} & ${\mathcal{O}}\left( {\frac{1}{{{{\xi}^3}}}} \right)$& \makecell[c]{Optimal\\ solution}\\
\hline
\centering
\makecell[c]{AO-NOMA} & \makecell[c]{Proposed \\ scheme} & \multicolumn{3}{c|}{\makecell[c]{Alternating \\ optimization}} & ${\mathcal{O}}\!\!\left( \!{N_s}\!{I_{ite}^{A}\!\!\left( \!\!{{{ {{\left( {3K\! +\! 4} \right)}} }^{3.5}} \!\!+\!\! I_{ite}^S\!\!\left( \!{\max \left(\! {M,\!2K\! -\! 1}\! \right)^4\!\sqrt M \!\log\! \frac{1}{{\epsilon}}}\! \right)}\! \right)}\! \right)$& \makecell[c]{Suboptimal\\ solution}\\
\hline
\centering
\makecell[c]{AO-FDMA} & \diagbox{\qquad \qquad}{\qquad \qquad} & \multicolumn{3}{c|}{\makecell[c]{Alternating \\ optimization}} & ${\mathcal{O}}\!\!\left(\!{N_s}\! {I_{ite}^{A}\!\!\left(\!\! {{{ {{\left( {2K \!+\! 4} \right)}} }^{3.5}} \!\!+\!\! I_{ite}^S\!\!\left( \!{\max \left( \!{M,\!K\! -\! 1} \!\right)^4\!\sqrt M \!\log\! \frac{1}{{\epsilon}}}\! \right)}\! \right)}\! \right)$& \makecell[c]{Suboptimal\\ solution}\\
\hline
\centering
\makecell[c]{AO-TDMA} & \diagbox{\qquad \qquad}{\qquad \qquad} &\diagbox{\qquad \qquad}{\qquad \qquad} &\multicolumn{2}{c|}{\makecell[c]{Alternating \\ optimization}} & ${\mathcal{O}}\left( {{{\left( {2K + 4} \right)}^{3.5}}} \right)$& \makecell[c]{Suboptimal\\ solution}\\
\hline
\end{tabular}}
\end{center}
\label{comparison}
\end{table*}
\indent Comparing the performance of the proposed algorithms, the solution obtained with the MO based algorithm serves as an upper bound for the optimal solution of the original problem, since SDR enlarges the feasible set of the IRS reflection coefficients. With the closed-form solutions for the optimal IRS reflection coefficients, ``EX-TDMA'' is capable of finding the global optimal solution. Furthermore, the AO based algorithms provide a suboptimal solution for the original problems. The tightness of the upper bound and the suboptimal solutions will be evaluated in Section VI.
\vspace{-0.8cm}
\section{Numerical Results}
\begin{figure}[t!]
\centering
\begin{minipage}[t]{0.45\linewidth}
\includegraphics[width=2.5in]{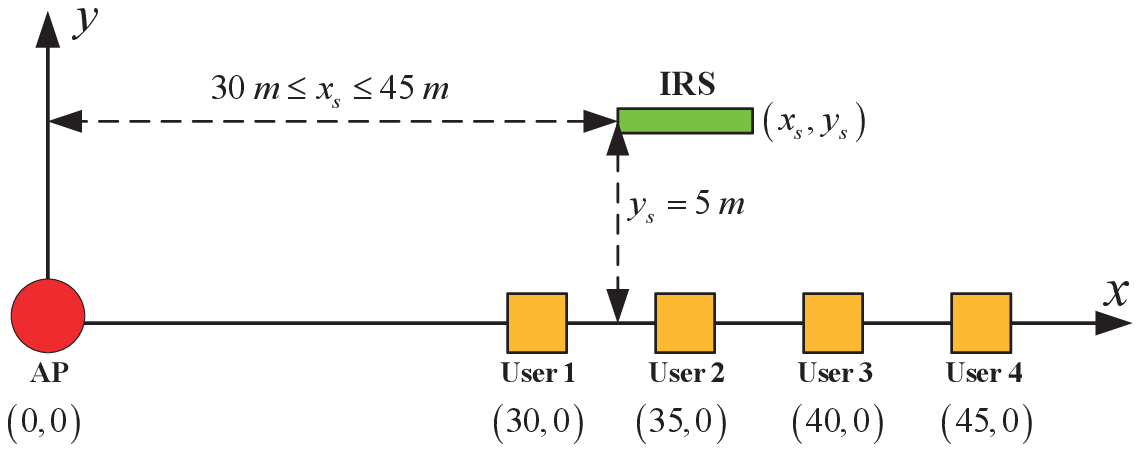}
\caption{Simulated IRS-assisted 4-user communication scenario (top view).}
\label{setup}
\end{minipage}
\quad
\begin{minipage}[t]{0.45\linewidth}
\includegraphics[width=2.5in]{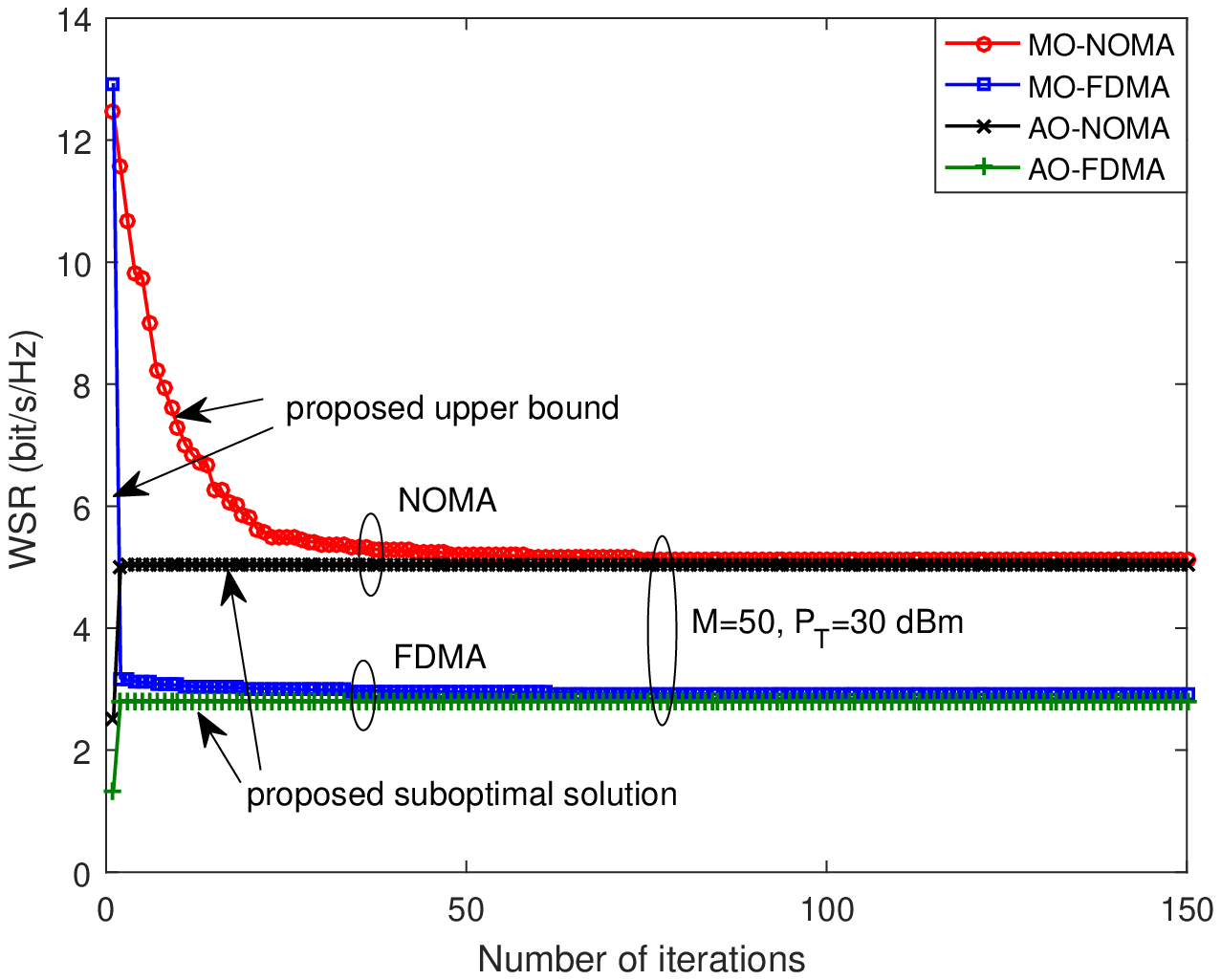}
\caption{Convergence of the proposed upper bound and suboptimal algorithms for NOMA and FDMA.}
\label{Convegency}
\end{minipage}
\end{figure}

\vspace{-0.2cm}
In this section, numerical results are provided to evaluate the effectiveness of proposed designs. We consider a scenario with $K=4$ users. The AP is located at ${\mathbf{b}} = {\left( {{0},{0},{5}} \right)^T}$ meters and the $k$th user is located at ${{\mathbf{u}}_k} = {\left( {{25+5k},{0},{1.5}} \right)^T}$ meters. The IRS deployment region is set as ${\mathbf{s}} \in \Omega  = \left\{ {\left( {{x_s},5,5} \right)^T|30 \le {x_s} \le 45} \right\}$. The horizontal locations of AP, IRS, and users are illustrated in Fig. \ref{setup}. The UPA at the IRS is parallel to the $x-z$ plane. For the number of IRS elements, we set ${M_h}=5$ and increase ${M_v}$ linearly with $M$. The element spacing is set to ${d_I} = \frac{\lambda }{2}$. The path loss exponents of the AP-IRS and IRS-user links are set as ${\alpha _{AI}} = {\alpha _{IU}} = 2.2$. The Rician factors of the AP-IRS and IRS-user links are set as ${\beta _{AI}} = {\beta _{IU}} = 3$ dB. We consider two different user rate weight vectors, namely, ${\mathbf{w}_1} = \left[ {0.1,0.2,0.3,0.4} \right]$ and ${\mathbf{w}_2} = \left[ {0.25,0.25,0.25,0.25} \right]$. For the other parameters, we set ${\rho _0} =  - 30$ dB and the noise power is ${\sigma ^2} =  - 90$ dBm. \textcolor{black}{All WSR results are obtained as follows: run the proposed algorithms once based on the LoS components to determine the IRS deployment location, and then again for each of the 100 independent channel realizations to obtain the average WSR.}
\vspace{-0.6cm}
\subsection{Selection of the Value of $\Delta$ for Local Region Optimization}
\vspace{-0.2cm}
For the considered simulation setup in Fig. \ref{setup}, to ensure that the AoAs/AoDs are approximately constant for our proposed local region optimization method, the value of $\Delta$ should satisfy the following condition
\vspace{-0.6cm}
\begin{align}\label{delta1}
\frac{\Delta }{{{y_s}}} \le {\varepsilon _{\max }}.
\end{align}
\vspace{-0.7cm}

\noindent Condition \eqref{delta1} means that the ratio of the maximum horizontal location change $\Delta$ along the $x$-axis and the horizontal distance between the IRS and the $x$-axis is below a threshold ${\varepsilon _{\max }}$. Therefore, the value of $\Delta$ can be obtained as $\Delta  \le {\varepsilon _{\max }}{y_s}$. Though very small values of ${\varepsilon _{\max }}$ increase the accuracy of the approximation, they also increase the number of iterations needed for convergence. To balance between accuracy and complexity, in this paper, the threshold is set to ${\varepsilon _{\max }} = 0.01$ and $\Delta=0.05$ meter. Therefore, the constraint in \eqref{delta} becomes $\left\| {{x_s} - x_s^{\left( l \right)}} \right\| \le 0.05$ for the considered simulation setup. Furthermore, the number of initialized candidate local points for the proposed AO algorithm is set to ${N_s} = 4$ and $\left\{ {{\mathbf{s}}_n^{\left( 0 \right)} = {{\left( {25 + 5n,5,5} \right)}^T}} \right\}$.
\vspace{-0.6cm}
\subsection{Convergence of Proposed Algorithms}
\vspace{-0.2cm}
\begin{figure}[b!]
\centering
\begin{minipage}[t]{0.45\linewidth}
\includegraphics[width=2.5in]{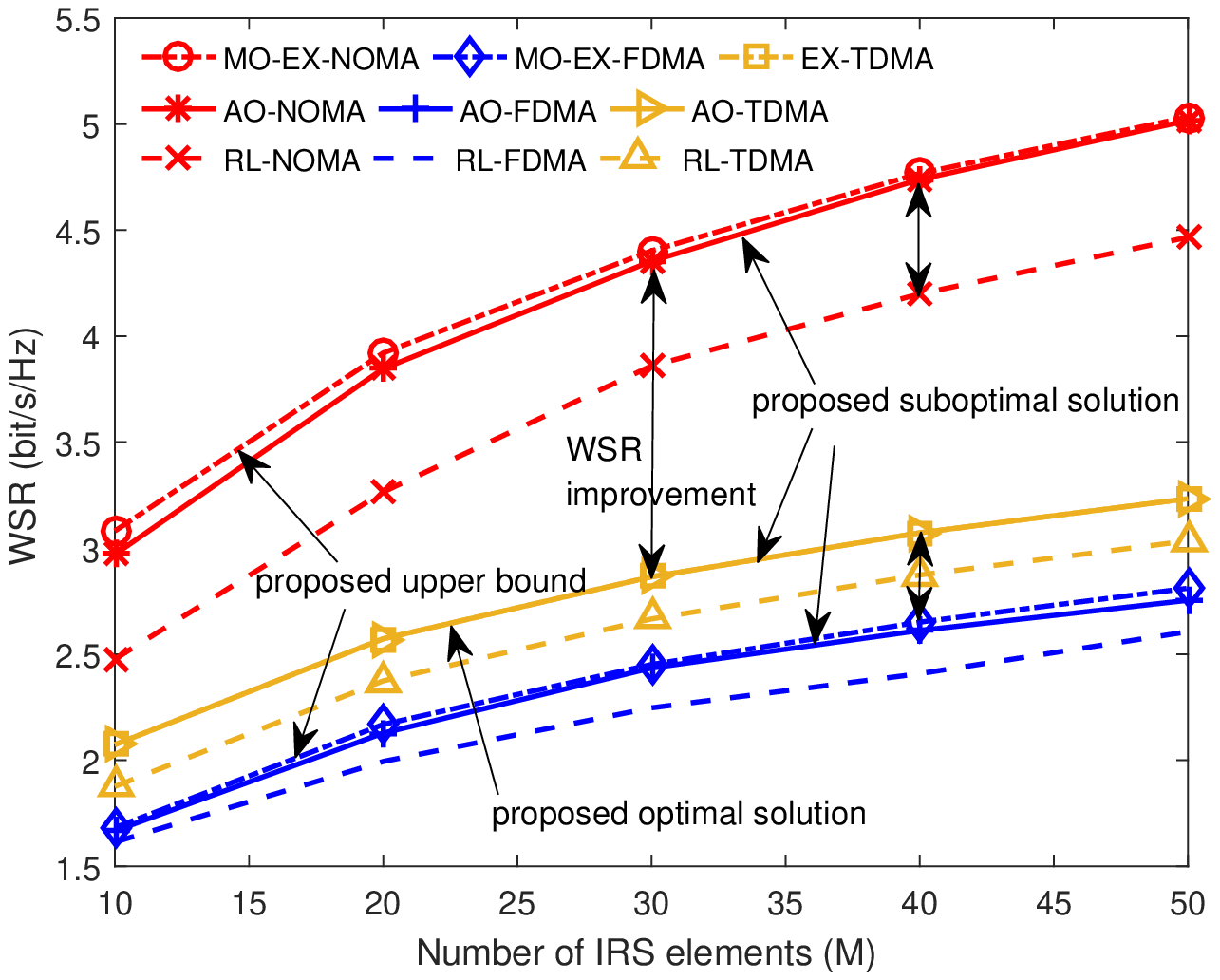}
\caption{WSR versus the number of IRS elements.}
\label{WSRvM}
\end{minipage}
\quad
\begin{minipage}[t]{0.45\linewidth}
\includegraphics[width=2.5in]{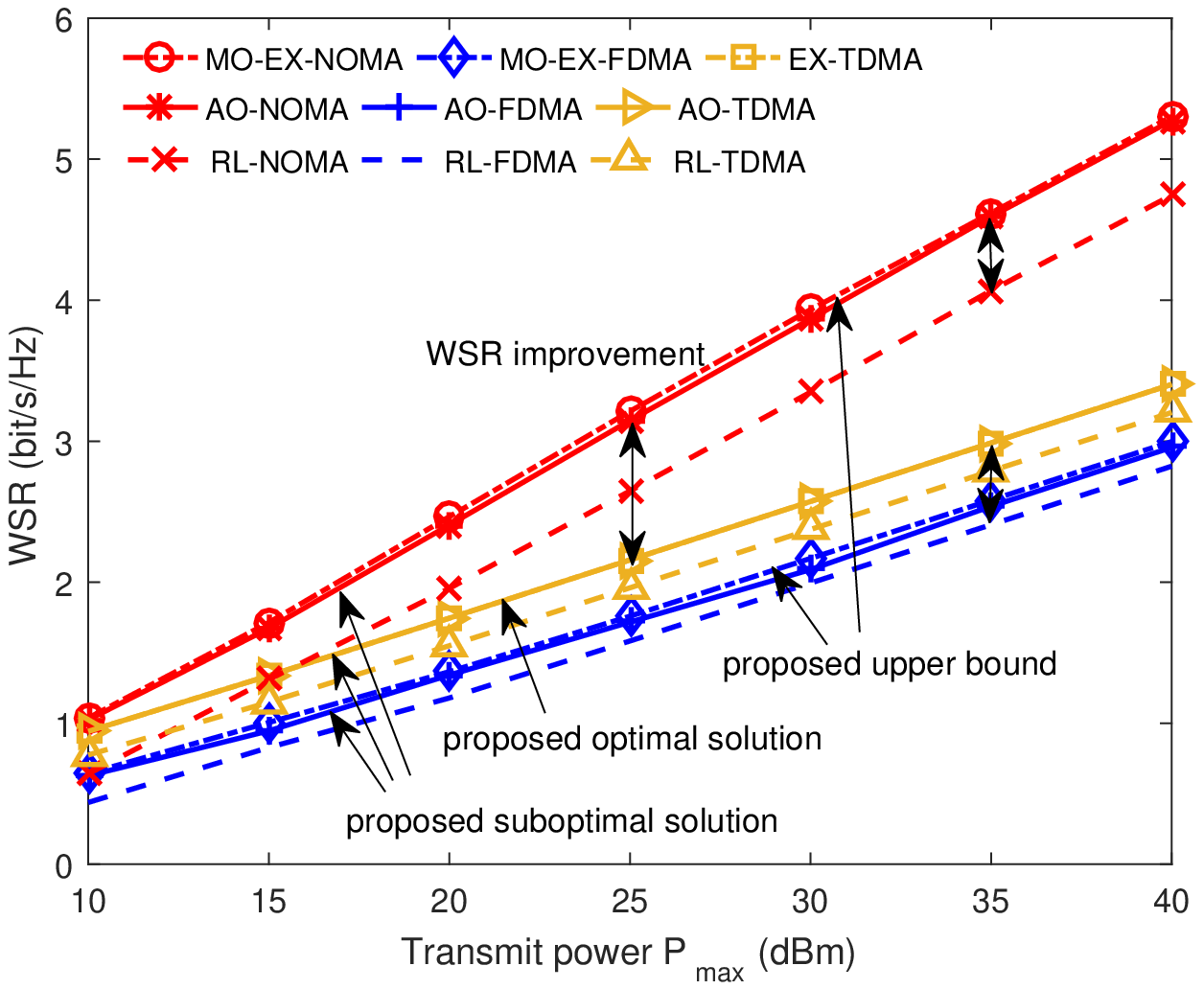}
\caption{WSR versus total transmit power.}
\label{WSRvP}
\end{minipage}
\end{figure}
Fig. \ref{Convegency} illustrates the convergence of the proposed upper bound and suboptimal algorithms for NOMA and FDMA for user rate weight vector ${{\mathbf{w}}_1}$. For a fair comparison, the convergence for both algorithms was studied for a given IRS deployment location and a given user decoding order. The maximum transmit power is set to ${P_{\max }} = 30$ dBm and the number of reflecting elements is set to $M=50$. As illustrated in Fig. \ref{Convegency}, the proposed upper bound and suboptimal algorithms converge as the number of iterations increase. Specifically, the proposed upper bound algorithm converges in less than 100 iterations and the proposed suboptimal algorithm converges to a similar value in less than 15 iterations, which is consistent with \textbf{Remark \ref{convegence}}.  The proposed suboptimal algorithm converges significantly faster than the proposed upper bound algorithm. It is also observed that the performance gap between the upper bound and the suboptimal solution is negligible, which illustrates the effectiveness of the proposed AO algorithms.
\vspace{-0.6cm}
\subsection{WSR versus $M$ and $P_{\max}$}
\vspace{-0.2cm}
In this subsection, we investigate the achieved WSR of the proposed algorithms. For comparison, we also consider the following benchmark scheme:
\vspace{-0.2cm}
\begin{itemize}
    \item \textbf{Random Location (RL)-X:} In this case, the IRS location is randomly selected within the feasible region. The other optimization variables are obtained with the proposed AO algorithms. ``X'' stands for the employed MA scheme.
\end{itemize}

\indent Fig. \ref{WSRvM} shows the WSR versus the number of IRS reflecting elements $M$ for $P_{\max}=30$ dBm and user rate weight vector ${{\mathbf{w}}_1}$. First, we observe that the achieved WSR of all schemes increases with $M$, since a larger number of IRS reflecting elements leads to higher passive array gains. In particular, NOMA has the best performance since all users can be served simultaneously in every time-frequency resource block. For the OMA schemes, TDMA achieves a higher WSR than FDMA due to the time selectivity of the IRS, which allows the users for TDMA to be served with the best channel power gains. Furthermore, it can be observed that the proposed suboptimal AO algorithms are capable of achieving near-optimal performance, closely approaching the proposed upper bound. Regarding the benchmark scheme, a considerable performance loss is observed for all three MA schemes, which underscores the importance of optimizing IRS deployment.\\
\indent Fig. \ref{WSRvP} shows the WSR versus transmit power $P_{\max}$ for $M=20$ and user rate weight vector ${{\mathbf{w}}_1}$. The WSR of all schemes increase as $P_{\max}$ increases and the performance gain of NOMA over OMA becomes more pronounced for larger transmit powers. The proposed suboptimal solutions cause a negligible performance loss compared to the proposed upper bound. The random IRS deployment locations lead to a worse performance compared to the proposed algorithms.
\begin{figure}[b!]
\centering
\begin{minipage}[t]{0.45\linewidth}
\includegraphics[width=2.5in]{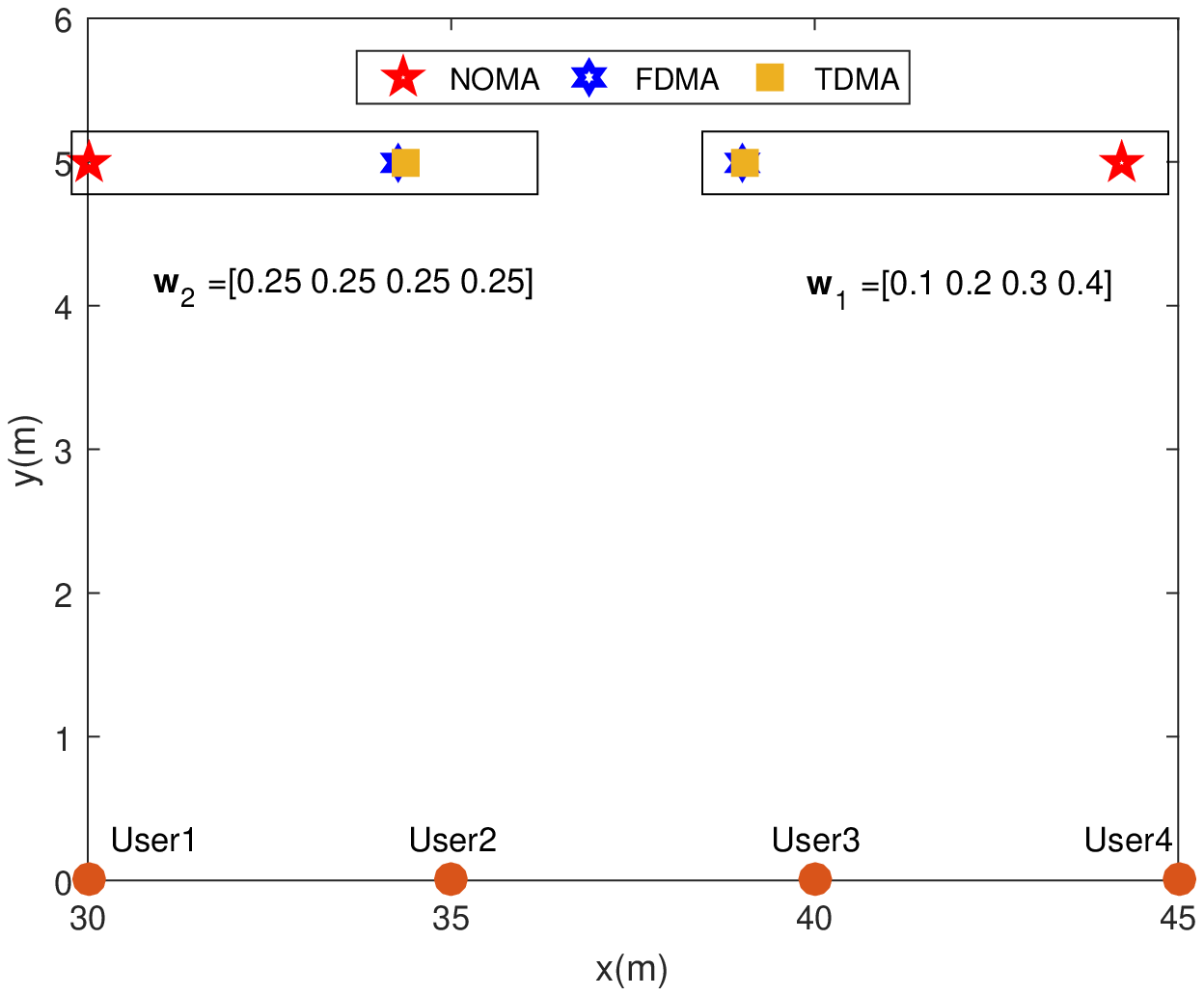}
\caption{Optimal IRS deployment location for different schemes and user weight vectors.}
\label{deploy}
\end{minipage}
\quad
\begin{minipage}[t]{0.45\linewidth}
\includegraphics[width=2.5in]{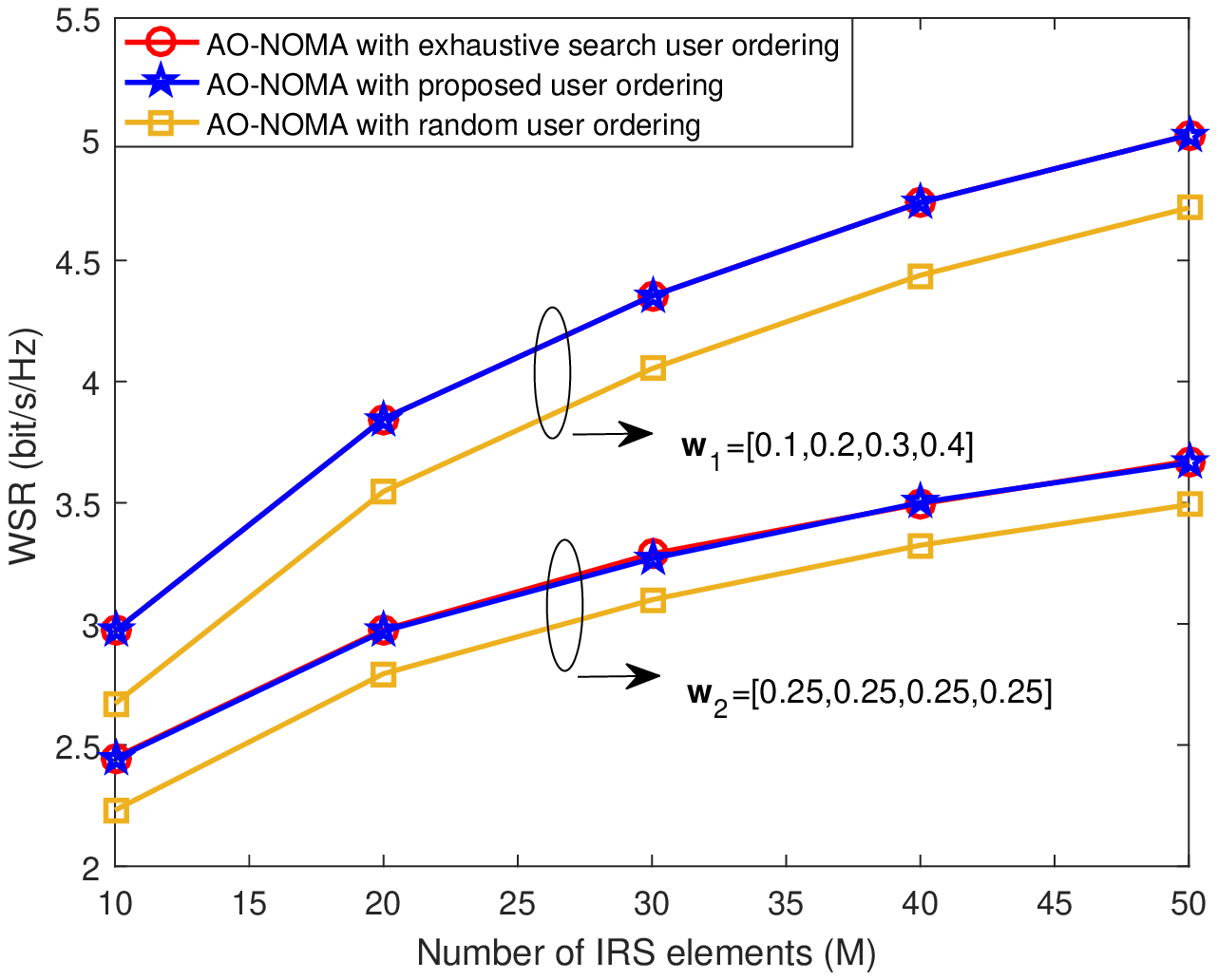}
\caption{WSR of different user ordering schemes with $P_{\max}=30$ dBm.}
\label{order compare}
\end{minipage}
\end{figure}
\vspace{-0.6cm}
\subsection{Optimal IRS Deployment Locations of Different Transmission Schemes}
\vspace{-0.2cm}
Fig. \ref{deploy} shows the optimal IRS deployment location obtained by the proposed AO algorithms for different MA schemes with user weight vetors ${{\mathbf{w}}_1}$ and ${{\mathbf{w}}_2}$. The transmit power is $P_{\max}=30$ dBm and $M=50$. For the case of ${\mathbf{w}_1} = \left[ {0.1,0.2,0.3,0.4} \right]$, the optimal IRS horizontal location for NOMA is $\left( {44.2,5} \right)$ meter, while the optimized IRS horizontal locations for TDMA and FDMA are almost the same at $\left( {39.3,5} \right)$ meter. Since users 3 and 4 have larger rate weights, the IRS for both NOMA and OMA is deployed to enhance the received signal strength of these users. However, the IRS deployment strategy for OMA is more \emph{symmetric} across all users than that for NOMA. This phenomenon can be explained based on the path loss model for IRS-assisted links, which is given by ${L_{IRS,k}} = \frac{{\rho _0^{}}}{{d_{AI}^{{\alpha _{AI}}}}}\frac{{\rho _0^{}}}{{d_{IU,k}^{{\alpha _{IU}}}}}$. Under the simulation setup considered in Fig. \ref{setup}, the channel power gain of user $k$ increases when the IRS gets close, and decreases when it moves away. Therefore, for OMA, the IRS is deployed to enhance the channel power gains of users 3 and 4 while keeping the channel power gains of users 1 and 2 at a moderately high level. However, for NOMA, the IRS is deployed to increase the channel power gain of user 4 and to decrease the channel power gains of the other users. By doing so, the channel conditions of the users become more distinctive, which is a preferable setting for NOMA transmission. For the case of ${\mathbf{w}_2} = \left[ {0.25,0.25,0.25,0.25} \right]$, the optimal IRS horizontal locations for NOMA, FDMA, and TDMA are $\left( {30,5} \right)$ meter, $\left( {34.3,5} \right)$ meter, and $\left( {34.7,5} \right)$ meter, respectively. Though each user has the same rate weight, the IRS for both NOMA and OMA is deployed closer to users 1 and 2 who are closer to the AP. The IRS deployment location for OMA enhances the channel power gains of users 2, 3, and 4 while slightly sacrificing some channel power gain of user 1. For NOMA, it is preferable to deploy the IRS in an \emph{asymmetric} manner to achieve distinct channel conditions for different users. The results in Fig. \ref{deploy} provide useful guidelines for IRS deployment for different MA schemes.
\vspace{-0.6cm}
\subsection{Impact of Decoding Order Design}
\vspace{-0.2cm}
In this subsection, we evaluate the performance achieved by the proposed user ordering scheme. For comparison, we also consider the performance of two benchmark schemes. For benchmark scheme 1, the optimal decoding order is obtained with the exhaustive search. For benchmark scheme 2, the decoding order of the users is selected randomly. The other optimization variables are optimized with the proposed AO algorithm. Fig. \ref{order compare} shows that the proposed user ordering scheme achieves almost the same performance as the exhaustive search scheme, which confirms the effectiveness of the proposed scheme. In addition, random user ordering suffers from a substantial performance loss as compared to the other two schemes, which demonstrates the importance of a careful user decoding order design for IRS-assisted NOMA transmission. \textcolor{black}{We can also observe that the WSR achieved with ${\mathbf{w}_1}$ is higher than that with ${\mathbf{w}_2}$. This is due to the fact that the communication rate of the strongest user contributes the most to the WSR for NOMA. Therefore, assigning the highest rate weight to the strongest user in general leads to a higher WSR.}
\vspace{-0.8cm}
\section{Conclusions}
\vspace{-0.2cm}
In this paper, the joint IRS deployment and MA design for downlink IRS-assisted multi-user networks was investigated. The IRS deployment location, the reflection coefficients of the IRS, and the power allocation at the AP were jointly optimized for maximization of the WSR for NOMA, FDMA, and TDMA. To solve the resulting non-convex problems, MO and AO based algorithms were developed to obtain a performance upper bound and high-quality suboptimal solutions. Our numerical results showed that the proposed suboptimal algorithms are capable of achieving near-optimal performance and that a significant performance gain can be achieved by optimizing the IRS deployment location. \textcolor{black}{Furthermore, our results also revealed that an asymmetric IRS deployment strategy is preferable for NOMA, while a symmetric IRS deployment strategy is superior for OMA. This insight provides useful guidelines for practical IRS implementation.}\\
\indent \textcolor{black}{In this work, the deployment of one IRS was considered to improve communication in a quasi-static scenario with spatially close users, which may be ineffective when the users are widely distributed and/or high-mobility. For such scenarios, the optimal deployment of multiple IRSs or mobile IRSs (e.g., based on IRSs mounted on intelligent unmanned vehicles) is a promising direction for future research.}
\vspace{-0.6cm}
\bibliographystyle{IEEEtran}
\bibliography{mybib}

\end{document}